\documentclass[twocolumn]{aastex62}
\usepackage{natbib}
\usepackage{threeparttable}
\usepackage{amssymb}
\usepackage{amsmath}
\usepackage{graphicx}
\usepackage{CJKutf8}
\usepackage{textcomp}
\hypersetup{citecolor=blue,linkcolor=blue}

\shorttitle{The LEGA-C of nature and nurture in stellar populations at $z\sim0.6-1.0$}
\shortauthors{Sobral et al.}

\begin{document}
\title{The LEGA-C of nature and nurture in stellar populations of galaxies at $z\sim0.6-1.0$: D$_n$4000 and H$\delta$ reveal different assembly histories for quiescent galaxies in different environments} 

\author[0000-0001-8823-4845]{David Sobral}
\affiliation{Department of Physics, Lancaster University, Lancaster, LA1 4YB, UK}
\affiliation{Departamento de F\'{i}sica, Faculdade de Ci\^{e}ncias, Universidade de Lisboa, Edif\'{i}cio C8, Campo Grande, PT1749-016 Lisbon, Portugal}

\author[0000-0002-5027-0135]{Arjen van der Wel}
\affiliation{Sterrenkundig Observatorium, Universiteit Gent, Krijgslaan 281 S9, B-9000 Gent, Belgium}

\author[0000-0001-5063-8254]{Rachel Bezanson}
\affiliation{Department of Physics and Astronomy and PITT PACC, University of Pittsburgh, Pittsburgh, PA 15260, USA}

\author[0000-0002-5564-9873]{Eric Bell}
\affiliation{Department of Astronomy, University of Michigan, 1085 South University Ave., Ann Arbor, MI 48109, USA}

\author[0000-0002-9330-9108]{Adam Muzzin}
\affiliation{Department of Physics and Astronomy, York University, 4700 Keele Street, Toronto, ON, M3J 1P3, Canada}

\author[0000-0003-2388-8172]{Francesco D'Eugenio}
\affiliation{Sterrenkundig Observatorium, Universiteit Gent, Krijgslaan 281 S9, B-9000 Gent, Belgium}

\author[0000-0003-4919-9017]{Behnam Darvish}
\affiliation{Cahill Center for Astrophysics, California Institute of Technology, 1216 East California Boulevard, Pasadena, CA 91125, USA}

\author[0000-0002-9656-1800]{Anna Gallazzi}
\affiliation{INAF-Osservatorio Astrofisico di Arcetri, Largo Enrico, Fermi 5, I-50125 Firenze, Italy}

\author[0000-0002-9665-0440]{Po-Feng Wu}
\affiliation{Institute of Astronomy \& Astrophysics, Academia Sinica, Taipei 10617, Taiwan}

\author[0000-0001-5937-4590]{Michael Maseda}
\affiliation{Department of Astronomy, University of Wisconsin-Madison, 475 N. Charter Street, Madison, WI 53706, USA}

\author[0000-0003-2871-127X]{Jorryt Matthee}
\affiliation{Department of Physics, ETH Z\"urich, Wolfgang-Pauli-Strasse 27, 8093 Z\"urich, Switzerland}

\author[0000-0002-0943-0694]{Ana Paulino-Afonso}
\affiliation{CENTRA, Instituto Superior Tecnico, Universidade de Lisboa, Av. Rovisco Pais 1, 1049-001 Lisboa, Portugal}

\author[0000-0001-5937-4590]{Caroline Straatman}
\affiliation{Sterrenkundig Observatorium, Universiteit Gent, Krijgslaan 281 S9, B-9000 Gent, Belgium}

\author[0000-0002-8282-9888]{Pieter G.~van Dokkum}
\affil{Astronomy Department, Yale University, 52 Hillhouse Ave, New Haven, CT 06511, USA}

\correspondingauthor{David Sobral}
\email{d.sobral@lancaster.ac.uk}

\begin{abstract}
\noindent Galaxy evolution is driven by a variety of physical processes which are predicted to proceed at different rates for different dark matter haloes and environments across cosmic times. A record of this evolution is preserved in galaxy stellar populations, which we can access using absorption-line spectroscopy. Here we explore the large LEGA-C survey (DR3) to investigate the role of the environment and stellar mass on stellar populations at $z\sim0.6-1$ in the COSMOS field. Leveraging the statistical power and depth of LEGA-C, we reveal significant gradients in D$_n$4000 and H$\delta$ equivalent widths (EWs) distributions over the stellar mass vs environment 2D spaces for the massive galaxy population ($M>10^{10}$\,M$_{\odot}$) at $z\sim0.6-1.0$. D$_n$4000 and H$\delta$ EWs primarily depend on stellar mass, but they also depend on environment at fixed stellar mass. By splitting the sample into centrals and satellites, and in terms of star-forming galaxies and quiescent galaxies, we reveal that the significant environmental trends of D$_n$4000 and H$\delta$ EW when controlling for stellar mass are driven by quiescent galaxies. Regardless of being centrals or satellites, star-forming galaxies reveal D$_n$4000 and H$\delta$ EWs which depend strongly on their stellar mass and are completely independent of the environment at $0.6<z<1.0$. The environmental trends seen for satellite galaxies are fully driven by the trends that hold only for quiescent galaxies, combined with the strong environmental dependency of the quiescent fraction at fixed stellar mass. Our results are consistent with recent predictions from simulations that point towards massive galaxies forming first in over-densities or the most compact dark matter haloes.
\end{abstract}

\keywords{galaxies: evolution --- galaxies: formation --- galaxies: stellar content}

\section{Introduction}

Our understanding of how galaxies form and evolve has changed dramatically over the last few decades \citep[e.g.][]{Dressler1980,Lilly96,MadauD2014}. This has been driven by an intense observational effort \citep[e.g.][]{Poggianti2006,Sobral2013,Bouwens2014,Darvish2016,PA2020}, along with modelling and theoretical breakthroughs, particularly due to state-of-the-art semi-analytical \citep[e.g.][]{Henriques2015} and hydrodynamical simulations \citep[e.g. Illustris, EAGLE, FIRE;][]{Hopkins2014FIRE,Vogelsberger2014,Genel2014,Schaye2015,Pillepich2018} and some direct comparisons with observations \citep[e.g.][]{Wu2021}.

In the local Universe, star formation and many galaxy properties depend strongly on the environment. Clusters are primarily populated by passive galaxies, while star-forming galaxies are mainly found in lower-density environments \citep[e.g.][]{Dressler1980}. It is also well-established \citep[e.g.][]{Best04} that the fraction of star-forming galaxies decreases with increasing local galaxy density (often projected local density, $\Sigma$) in the local Universe, at low \citep[$z<0.5$, e.g.][]{Kodama2004} and at intermediate \citep[$z\sim1$, e.g.][]{Poggianti2006,Sobral11,Woo2013} redshifts. Stellar mass and internal processes connected with the stellar mass assembly also play a key role, with stellar mass becoming an even better predictor of galaxy properties irrespective of environment towards higher redshift \citep[e.g.][]{Woo2013,Darvish2016,Darvish2017,Webb2020}. While mass and local environmental density typically correlate, it is now possible to disentangle their roles in the local Universe and at higher redshift and to show that both are relevant for quenching star-formation \citep[e.g.][]{Peng10,Sobral11,Darvish2016}. Furthermore, recent studies have also investigated the role of the different components of the ``cosmic web" \citep[e.g.][]{Darvish2017,Tojeiro2017,Goh2019}, although the majority find that environmental trends are primarily driven by galaxy environmental density and only mildly depend on the component of the cosmic web, such as fields, filaments and clusters.

Several studies in a broad range of redshifts have shown that, on average, many properties of star-forming galaxies that are directly or indirectly linked to star formation activity (e.g. star formation rates, specific star formation rates, emission line equivalent widths, main-sequence of star-forming galaxies) seem to be invariant to their environment \citep[e.g.][]{Peng10,Wijesinghe12,Muzzin12,Koyama13,Koyama14,Hayashi14,Darvish14,Darvish15,Darvish2016}. Therefore, the main role of the environment might be to set the fraction of quiescent/star-forming galaxies \citep[e.g.][]{Peng10,Sobral11,Muzzin12,Darvish14,Darvish2016,Old2020}; but how quickly and through which mechanisms does that happen? In addition, not all characteristics of star-forming galaxies are independent of environment. For example, metallicities and electron densities have been shown to be a function of environment \citep[e.g.][]{Kulas2013,Shimakawa15,Harshan2020}, with studies finding that star-forming galaxies have lower electron densities, slightly higher metallicities and also higher dust extinction at fixed stellar mass \citep[e.g. higher $A_{\rm H\alpha}$;][]{Sobral2016E} in high-density environments when compared to lower density/more typical environments \citep[][]{Sobral2015,Darvish15}. There is also evidence in favour of the environment being crucial in setting the amount of rotational support \citep[e.g.][]{Cappellari2011,Eugenio2013,Cole2020}, but see also \cite{Brough2017} and \cite{Veale2017}.

Finding the exact mechanisms of galaxy quenching and their physical agents is still one of the unsolved problems in galaxy evolution. Many internal and external processes are thought to be linked to the quenching process. These include ram pressure stripping \citep[e.g.][]{Gunn72}, strangulation \citep[e.g.][]{Larson80,Balogh00}, galaxy-galaxy interactions and harassment \citep[e.g.][]{Mihos96,Moore98}, tidal interaction between the potential well of the environment and the galaxy \citep[e.g.][]{Merritt84,Fujita98}, or halo quenching \citep[e.g.][]{Birnboim03,Dekel06}. The strength and quenching timescale of each physical process vary depending on many parameters such as the properties of the quenching environment (e.g. density, temperature, velocity dispersion, dynamical state) and those of the galaxy being quenched (e.g. mass, satellite vs. central). Those processes should, in principle, leave an imprint in the star formation history of galaxies in different environments.

There is substantial evidence for the truncation of the atomic and molecular gas content of cluster galaxies, possibly due to ram pressure stripping (e.g. \citealp{Cayatte90,Boselli08,Fumagalli09}; see also \citealp{Boselli14}). In principle, the dust distribution within galaxies should also depend on their host environment, most likely truncated outside-in as galaxies fall into the deeper potential wells of denser clusters \citep[e.g.][]{Noble2017,Zavala2019}.
 
One might naively expect a continuous decline in the star formation of galaxies from the field to the dense cores of clusters, which would be reflected in observables such as nebular emission lines, Balmer absorption lines or the strength of the 4000\,{\AA} break (D$_{4000}$). However, before galaxies undergo a full quenching process in dense regions, they may experience a temporary enhancement in star formation activity, which may complicate how observations are interpreted. For example, ram pressure stripping can initially compress the gas/dust which is favourable for star formation, and thus increasing the column density of the gas and dust \citep[e.g.][]{Gallazzi09,Bekki2009,Owers2012,Roediger2014}. Tidal galaxy-galaxy interactions can lead to the compression and inflow of the gas in the periphery of galaxies into the central part, feeding and rejuvenating the nuclear activity, which results in a temporary enhancement in star formation activity \citep[e.g.][]{Mihos96,Kewley06,Ellison08}. Intermediate-density environments such as galaxy groups, in-falling regions of clusters, cluster outskirts, merging clusters and galaxy filaments provide the ideal conditions for such interactions \citep[e.g.][]{Moss06,Perez09,Tonnesen12,Stroe2014,Stroe2015,Stroe2020,Stroe2021}. Therefore, one might expect a temporary enhancement in star formation activity and dusty star formation in intermediate-density environments before the galaxies quench. This has been found in several studies: intermediate-density environments seem to be sites of enhanced star formation rate, star-forming fraction and obscured star formation activity \citep[e.g.][]{Smail99,Best04,Koyama08,Gallazzi09,Koyama10,Sobral11,Coppin12}. In practice, filaments may well be the actual dominant intermediate-density environment \citep[e.g.][]{Darvish14,Darvish15}.

Until now, most detailed studies were only possible for the local Universe \citep[see][and references therein]{Gallazzi2021} and/or for specifically targeted clusters \citep[e.g.][]{Muzzin12,Balogh2021}. With the availability of LEGA-C \citep[][]{vanderWel2016}, it is now possible to look for small absolute differences in the stellar populations directly and not just from recombination lines and/or photometry. LEGA-C allows us to probe to a look-back time ($\sim7$\,Gyrs ago) when galaxies have much younger stellar populations than the local Universe and when the roles of stellar mass and environment might be significantly different. In this paper we study how key galaxy spectral properties that are correlated with star formation history depend on both stellar mass and on local environmental density \citep[see also][]{Bezanson2018b,Wu2018,Wu2018b}. This paper is a first attempt at separating the underlying effects of ``nature" and "nurture" on stellar populations and star formation histories \citep[see also][]{Chauke2018} at $z\sim0.6-1.0$ by leveraging the large sample size and exquisite environmental range of LEGA-C. Section \ref{Sample} briefly presents the sample, the measurements and the methods. In Section \ref{results} we show the results and then discuss them in Section \ref{Discussion}. Finally, Section \ref{conclusion} presents the conclusions. We use AB magnitudes \citep[][]{Oke1983}, a \cite{Chabrier2003} initial mass function (IMF) and assume a cosmology with H$_{0}$=70\,kms$^{-1}$Mpc$^{-1}$, $\Omega_{M}$=0.3 and $\Omega_{\Lambda}$=0.7.

\section{The LEGA-C sample at $z\sim0.6-1.0$}\label{Sample}

\subsection{The LEGA-C survey and the parent sample}

LEGA-C, the Large Early Galaxy Astrophysics Census \citep[][]{vanderWel2016,vanderWel2021,Straatman2018}, used the Visible Multi-Object Spectrograph (VIMOS; \citealt{LeFevre2003}) mounted on the 8\,m Very Large Telescope (VLT) to obtain rest-frame optical spectra of $\sim 3000$ $K_s$-band selected galaxies mainly at $0.6 \leq z \leq 1.0$. Each target was observed for $\sim 20$\,hours at a spectral resolution of $R\sim3500$, leading to a typical continuum signal-to-noise ratio (S/N) of 20 \AA$^{-1}$, unprecedented for a large statistical sample at $z\sim1$.

\begin{figure}
 \centering
 \includegraphics[width=8.5cm]{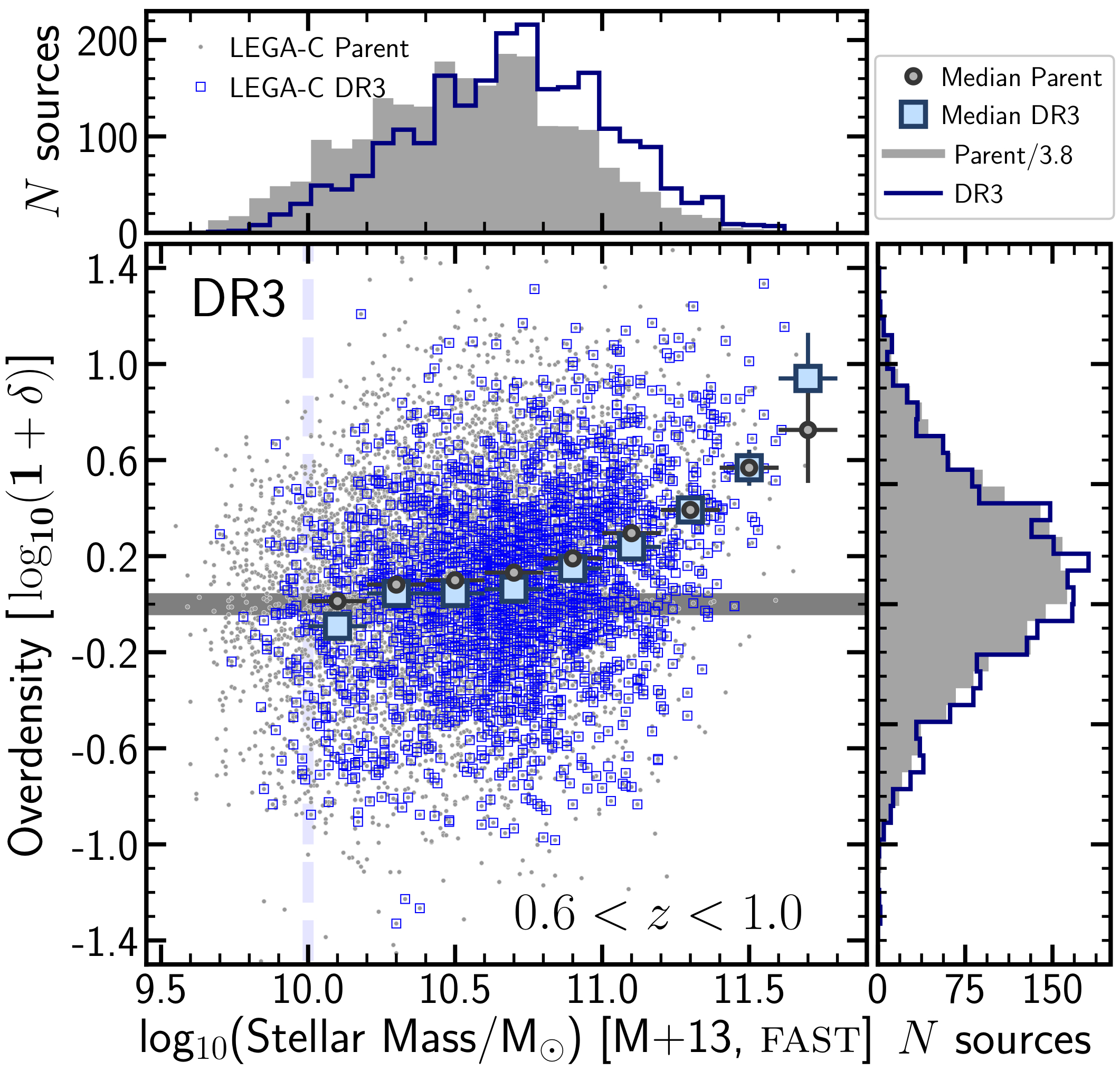}
 \caption{The relation between stellar mass and over-density ($\log_{10}(1+\delta)$) for the full DR3 in LEGA-C and the LEGA-C parent sample based on \citet{Muzzin2013}. In order to directly compare the parent sample with DR3 we use stellar masses presented in \citet{Muzzin2013} which use {\sc fast}. The parent sample (number) distribution in stellar mass and over-density has been divided by 3.8 so it can be directly compared with DR3. The differences in the stellar mass distribution can be very well corrected using LEGA-Cs completeness correction. We find that LEGA-C DR3 samples the overall range of environments of the population of galaxies. We note that some minor number of lower mass galaxies are at the highest over-densities that get preferentially missed as LEGA-C prioritises higher mass galaxies.}
 \label{Overdensity_distribution_parent}
\end{figure}

We use data from LEGA-C DR3 \citep[][]{vanderWel2021}. LEGA-C's primary (parent) sample consists of galaxies brighter than $K_s = 20.7 - 7.5 \times \log((1+z)/1.8)$ and with redshifts $0.6 \leq z \leq 1.0$ \citep{vanderWel2016}. The high-quality LEGA-C spectra allows us to accurately measure D$_n$4000 and H$\delta$ (stellar, in absorption, subtracting any emission-lines), along with many other stellar features (see \citealt{vanderWel2021}). Stellar masses for the parent and previous data releases of LEGA-C have been computed using {\sc fast} \citep[][]{Kriek2009} and all the available photometry in the COSMOS field. The {\sc fast} stellar mass distribution of the full DR3 LEGA-C sample and that of the COSMOS parent sample used to allocate slits is shown in Figure \ref{Overdensity_distribution_parent}. LEGA-C is (by design) more complete at higher masses than at low masses \citep[see][including for more information on the robust completeness corrections.]{Straatman2018}.

\begin{figure}
 \centering
 \includegraphics[width=8.3cm]{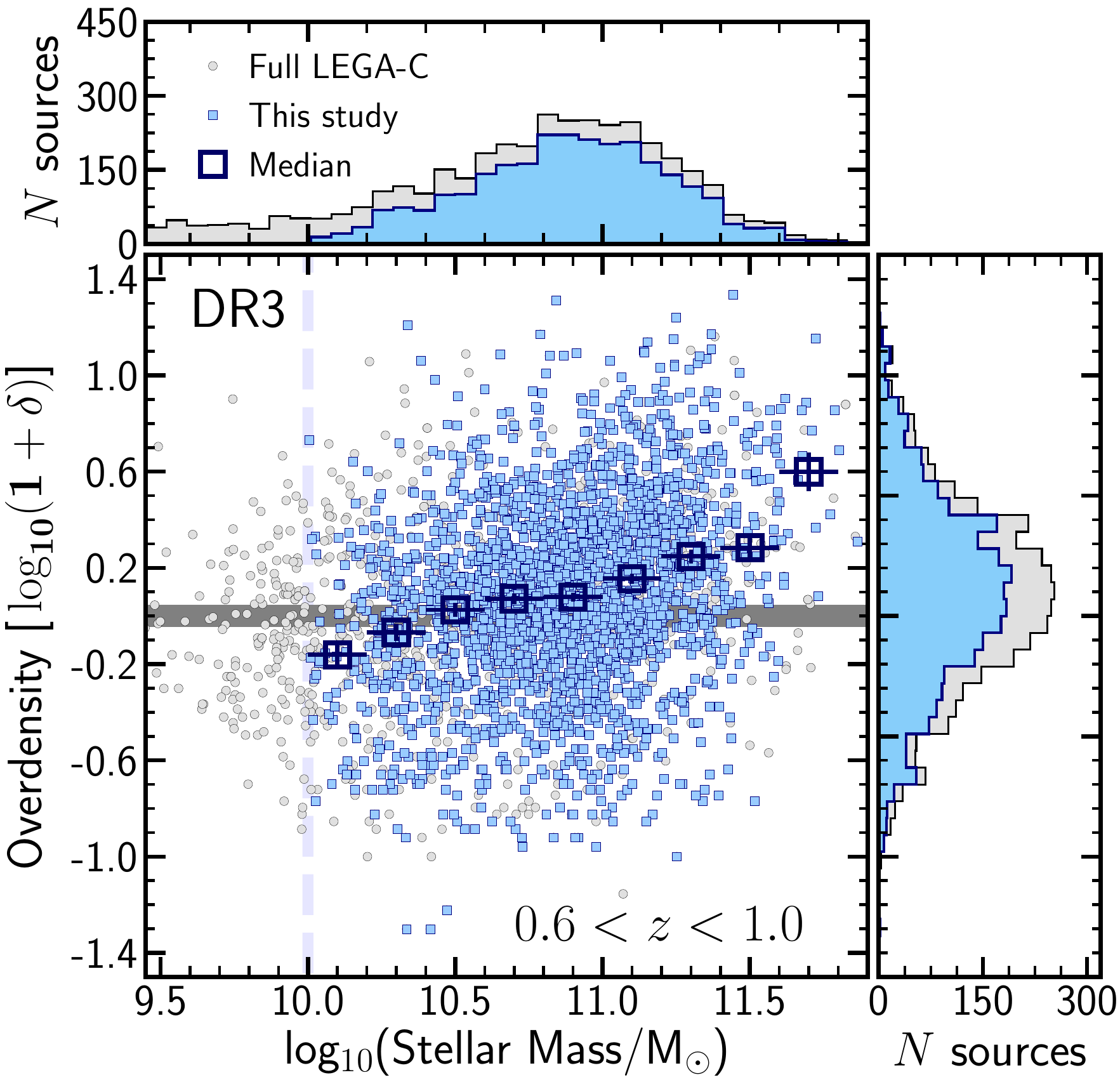}
 \caption{The relation between stellar mass ({\sc Prospector}-derived) and over-density ($\log_{10}(1+\delta)$) for the full DR3 in LEGA-C and for the sample used in the paper. We find a statistically significant correlation between stellar mass and median environmental density with $\log_{10}(1+\delta)=(0.38\pm0.04)\log_{10}[M/M_\odot]-(4.1\pm0.4)$. Galaxies with 10$^{10}$\,M$_{\odot}$ typically reside in slightly under-dense regions at $z\sim0.6-1.0$, while galaxies with $>10^{11}$\,M$_{\odot}$ typically reside in regions that are $\approx2.5\times$ over-dense, but note the significant scatter. This trend is primarily driven by quiescent galaxies.}
 \label{Overdensity_distribution}
\end{figure}

For most of the analysis in this paper we use and explore new {\sc Prospector} \citep[][]{Leja2019} stellar masses \citep[see][for details]{vanderWel2021}, which are typically higher than {\sc fast} stellar masses, due to older ages and different star formation histories. Figures \ref{Overdensity_distribution_parent} and \ref{Overdensity_distribution} allow a direct visualisation of the small differences between {\sc fast} (Figure \ref{Overdensity_distribution_parent}) and {\sc Prospector} (Figure \ref{Overdensity_distribution}).

%%%%%%%%%%%%%%%%%%%%%%%%%%%%%%%%%%%%
%
%  Figure 1 - Mass vs Environment: distribution of each and the correlation between both
%
%%%%%%%%%%%%%%%%%%%%%%%%%%%%%%%%%%%%

%%%%%%%%%%%%%%%%%%%%%%%%%%%%%%%%%%%%
%
%  Figure 2 - Mass vs Environment: distribution of each and the correlation between both
%
%%%%%%%%%%%%%%%%%%%%%%%%%%%%%%%%%%%%

\subsection{Environmental densities}\label{over_dens_calc}

In order to evaluate the environment in which each of our galaxies resides in, we use local over-densities estimated by \cite{Darvish14,Darvish2016,Darvish2017} over the full COSMOS field. Local over-densities are computed based on the full photometric redshift catalog by \cite{Ilbert2013}, which has the same selection as the LEGA-C parent sample \citep[][]{Muzzin2013}, over the full 1.8\,deg$^2$ COSMOS field. The full density field is calculated for  $0.05 < z < 3.2$ galaxies with $K_{s} < 24$. This enables us to probe the full cosmic web in COSMOS to a much lower stellar mass ($>10^{9.6}$\,M$_\odot$) than the completeness limit of LEGA-C. In this study we shall refer to over-density or local (galaxy) over-density as the normalised density, $1+\delta$, given by:

\begin{equation}
1+\delta= \frac{\Sigma}{\Sigma_{\rm median}},
\end{equation}
where $\Sigma_{\rm median}$ is the median of the density field of the redshift slice \citep[see][for full details]{Darvish2016,Darvish2017} each galaxy is in and $\Sigma$ is the absolute density of sources at the position of the source. For further details, and a comprehensive investigation of different methods to compute local over-densities, see \cite{Darvish2016,Darvish2017}.

%%%%%%%%%%%%%%%%%%%%%%%%%%%%%%%%%%%%
%
%  Figure 3 - 3D distribution: Redshift vs RA and Dec + Density
%
%%%%%%%%%%%%%%%%%%%%%%%%%%%%%%%%%%%%
\begin{figure*}
 \centering
 \includegraphics[width=14cm]{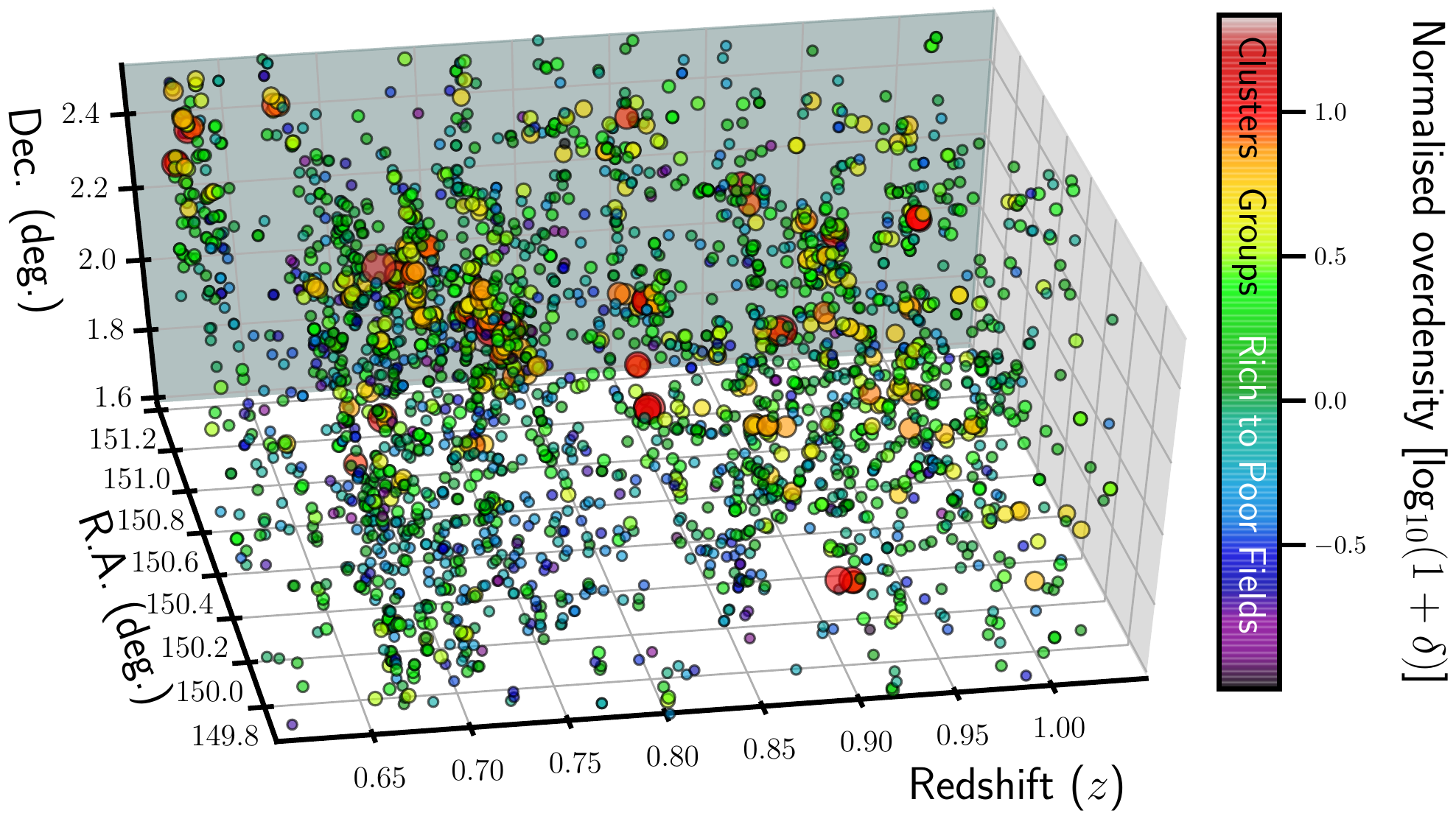}
 \caption{The 3D distribution of the LEGA-C sample used in this paper ($0.6<z<1.0$) in 3D, colour-coded by local over-density (see Section \ref{over_dens_calc}); symbol sizes also increase with increased over-density. Our sample traces the full range of densities available in the full COSMOS field, from relatively poor fields/under-dense regions to the cores of clusters. Note the large scale/filamentary structure around a massive structure/wall at $z\approx0.73$ \citep[see e.g.][]{Iovino2016}, and other rich groups at a variety of redshifts.}
 \label{3D_distribution}
\end{figure*}

We match each LEGA-C galaxy with the density field they are embedded in within the full COSMOS field, thus providing each galaxy with a value of $\log_{10}(1+\delta)$. We show the full distribution for our sources and also for the full LEGA-C sample in Figure \ref{Overdensity_distribution}. In Figure \ref{Overdensity_distribution_parent} we reveal the full distribution of stellar masses and environmental densities for LEGA-C's parent sample and for DR3. The spectroscopic LEGA-C DR3 sample extends over the overall range of environments of the general population of galaxies in the COSMOS field at $0.6 \leq z \leq 1.0$. We note that some minor number of lower mass galaxies at the highest over-densities are preferentially missed (compared to the full parent sample), as LEGA-C prioritises higher mass galaxies. Furthermore, one can see in the top histogram of Figure \ref{Overdensity_distribution_parent} that even after scaling, the distributions of LEGA-C DR3 and the parent sample are not the same, but the histogram on the right reveals that the environmental distributions are very well matched. Therefore, Figure \ref{Overdensity_distribution_parent} shows that completeness corrections are necessary as a function of mass, yet the sample is a very good representation of all environments and reveals a very similar relation between stellar mass and over-density as the parent sample.

%%%%%%%%%%%%%%%%%%%% TABLE  1   %%%%%%%%%%%%%%%%%%%%%%
%
%         Table with steps, flags applied, numbers averag properties and range in mass, density
%
%%%%%%%%%%%%%%%%%%%%%%%%%%%%%%%%%%%%%%%%%%%%%%%%%
\begin{table*}
\centering
\caption{Overview of the full DR3 LEGA-C sample \citep[see][]{Straatman2018,vanderWel2021}, together with all the flags, steps and cuts used to produce the final sample. Values are averages for the sub-samples and the corresponding standard deviation as the error.}
\begin{tabular}{lccccl}
\hline
Sample & Mass   &  $\log_{10}(1+\delta)$  & $z_{\rm spec}$ & \# sources  & Flags/cuts used for sub-sample     \\ \hline
(1) Full DR3   &  $10.7\pm0.5$   &  $0.1\pm0.4$  & $0.82\pm0.17$   & 4081  & ---  \\ 
(2) Primary  &   $10.9\pm0.4$   &  $0.1\pm0.4$  & $0.78\pm0.12$   & 2802  &  (1) \& flags=0 \& primary=1  \\ 
(3) $z$ cut     &  $10.9\pm0.4$   &  $0.1\pm0.4$  & $0.78\pm0.11$   & 2686   &   (2) \& $0.6\leq z \leq 1.0$  \\ 
(4) Mass cut  &  $10.9\pm0.3$   &  $0.1\pm0.4$  & $0.79\pm0.11$   & 2645  &   (3) \& M$>10^{10}$\,M$_{\odot}$     \\ 
(5) $\log_{10}(1+\delta)$  &  $10.9\pm0.3$   &  $0.1\pm0.4$  & $0.78\pm0.11$   & 2499  &   (4) \& $\log_{10}(1+\delta)$ measured    \\ 
(6) This study  &  $10.9\pm0.3$   &  $0.1\pm0.4$  & $0.81\pm0.11$   & 1900  &   (5) \& D$_n$4000 H$\delta$ measured     \\ 
\hline
\label{table_SAMPLE}
\end{tabular}
\end{table*}

Overall, we find that our sources span the full range of over-densities found in the COSMOS field: from poor fields to rich clusters\footnote{We note that we miss some of the most extreme under-densities in the COSMOS field, but we focus on densities from fields to clusters and not voids. This is not surprising given i) the stellar mass selection and ii) prioritising the more massive/brighter $K$-band sources.}, thus allowing us to investigate the potential role of the environment at $z\sim0.6-1.0$ with a single outstanding data-set.

In order to visualise the structures that correspond to a variety of densities, we show the on-sky distribution of our sample de-projected using the spectroscopic redshifts and labelled by local over-density in Figure \ref{3D_distribution}. We can see a large, over-dense region around $z\approx0.7$, a super-structure/cluster \citep[][]{Knobel2009,Iovino2016} containing an X-ray cluster, several structures with diffuse X-ray emission from its multi components and a large number of likely in-falling groups. Furthermore, high-density peaks are also found in rich groups/clusters at $z\approx0.68$, $z\approx0.7$ and $z\approx0.89$ within DR3 (see Figure \ref{3D_distribution}), along with sampling some parts of the $z\sim0.85$ COSMOS super-cluster \citep[see][]{PA2018,PA2019,PA2020}; all of these are also detected in the X-rays.

\subsubsection{The sample used in this paper}

From the 4081 sources in the full DR3 catalogue \citep[see also DR2;][]{Straatman2018} we apply different cuts using flags and other properties to define our final sample used in this study, as further detailed in Table \ref{table_SAMPLE}. Specifically, we apply all quality flags as described by \cite{Straatman2018}, a redshift cut of $0.6\leq z\leq 1.0$, a mass cut of $ \log_{10}(M_\ast/M_\odot) \geq 10.0$, we require environmental densities to be available, and we require that both D$_n$4000 and H$\delta$ measurements are reliable. At $z\sim0.8$, essentially all LEGA-C galaxies have both D$_n$4000 and H$\delta$ measurements, but the fraction of galaxies with both measurements drops to $\sim$60\% for $z\sim0.6$. This leads to a spectroscopic sample of 2499 galaxies and line indices spectroscopic sample of 1900 galaxies (see Table \ref{table_SAMPLE}).

%%%%%%%%%%%%%%%%%%%% TABLE 2   %%%%%%%%%%%%%%%%%%%%%%
%
%         Table with steps, flags applied, numbers average properties and range in mass, density
%
%%%%%%%%%%%%%%%%%%%%%%%%%%%%%%%%%%%%%%%%%%%%%%%%%
\begin{table*}
\centering
\caption{Overview of the number of sources in the LEGA-C sample used in terms of the cosmic web environment and group membership \citep[based on][]{Darvish2017}. All sources used in this study have an environmental density, group membership status and a likely cosmic web allocation. Note that for our analysis, ``isolated" galaxies will be combined with centrals.}
\begin{tabular}{ccccc}
\hline
Environment & \# of Centrals   &  \# of Satellites  &  \# of  Isolated &  \# Full sample    \\ 
(Cosmic web)     &  All [SF] (Q)   & All   [SF] (Q)  & All  [SF] (Q)  & All  [SF] (Q)   \\ \hline
Field & 339 [170] (169)  & 168 [112] (56)  & 473 [319] (154)  & 980 [601] (379)  \\ 
Filament & 257 [108] (149)  & 336 [198] (138)  & 148 [80] (68)  & 741 [386] (355)  \\ 
Cluster & 56 [9] (47)  & 114 [44] (70)  & 9 [4] (5)  & 179 [57] (122)  \\ 
Full web & 652 [287] (365)  & 618 [354] (264)  & 630 [403] (227)  & 1900 [1044] (856)  \\ 
\hline
\label{table_SAMPLE_env}
\end{tabular}
\end{table*}

The lower mass limit ensures that the $K_s$-band magnitude-limit of the LEGA-C survey does not introduce a strong bias against red galaxies at the lower mass end. Furthermore, we also apply corrections to bring the sample to a roughly mass-complete sample as detailed in \cite{Straatman2018}. Each LEGA-C galaxy has a specific correction factor assigned to it based on how complete it is compared to the parent mass complete sample (see also Figure \ref{Overdensity_distribution_parent}). These corrections are relatively small (factors $\sim 2-3$) above $\approx10^{10.5}$\,M$_{\odot}$, but they become significantly larger for lower masses close to $\sim10^{10}$\,M$_{\odot}$.

\subsection{Cosmic web, centrals, satellites and isolated LEGA-C galaxies}\label{CW_m_c}

In addition to allocating a local over-density value to each galaxy (see Section \ref{over_dens_calc}), we also assign our LEGA-C galaxies to different components of the cosmic web (Table \ref{table_SAMPLE_env}). The assignment is done depending on whether they are likely to reside in field, filaments or groups/clusters (based on the geometry of the cosmic web at the 3D position of each galaxy), together with determining whether they are centrals, satellites or isolated galaxies. In order to do this, we explore the public cosmic web catalogue provided by \cite{Darvish2017}, and we refer readers to that study for further details. Briefly, the density field is constructed with a mass-complete sample in the COSMOS field up to $z\sim1.2$, thus covering our full LEGA-C sample. Weighted adaptive kernel smoothing is used, with a global smoothing of 0.5\,Mpc, the typical virial radius for X-ray groups and clusters in the COSMOS field. The 2D version of the Multi-Scale Morphology Filter algorithm \citep[see][]{Darvish14} is used to compute a filament and cluster signal (between 0 and 1) depending on the resemblance of the density field in the position of a galaxy to either a filament (linear geometry) or a cluster (spherical geometry/distribution). These are used to assign galaxies to their position in the cosmic web, as fully detailed in \cite{Darvish2017}. In addition, we also explore the group catalogue and the classification of each LEGA-C galaxy provided by \cite{Darvish2017} to assign each LEGA-C galaxy as a central, satellite or isolated galaxy. Note that at the median redshift of LEGA-C, $z\approx0.85$, a satellite may have properties very similar to a central as they have not had time to be a satellite for long, unlike in the local Universe, where typically satellites have been satellites for more than 1\,Gyr. Table \ref{table_SAMPLE_env} provides an overview of the number of sources in each component of the cosmic web, and also the number that are classified as centrals, satellites and isolated galaxies. We will use these classifications in order to further explore trends with local density and stellar mass.

\subsection{The relation between stellar mass and environment}\label{mass_environment_}

As more massive galaxies are typically more biased and tend to reside in more massive dark matter haloes, it is expected that they will reside in higher densities. In Figure \ref{Overdensity_distribution} we show that there is a relation between stellar mass and local environment for our LEGA-C galaxies. Although there is a significant scatter, we find a statistically significant ($\approx10$\,$\sigma$) correlation between the two, which can be parameterised\footnote{We fit the median over-density as a function of stellar mass bins as shown in Figure \ref{Overdensity_distribution}. We perturb individual bins within their uncertainties and re-fit the data 10,000 times.} by $\log_{10}(1+\delta)=(0.43\pm0.04)\log_{10}[M/M_\odot]-(4.4\pm0.4)$. The relation between stellar mass and over-density implies that galaxies with 10$^{10}$\,M$_{\odot}$ typically reside in average densities at $z\sim0.6-1.0$ (Figure \ref{Overdensity_distribution}), while galaxies with $>10^{11}$\,M$_{\odot}$ typically reside in regions that are $\approx2.5\times$ over-dense, but note the significant scatter (see Figure \ref{Overdensity_distribution}), which is $\approx0.4$\,dex.

\subsection{Splitting the population into quiescent and star-forming: $UVJ$}

We classify galaxies as quenched or star-forming based on the $UVJ$ colours \citep[e.g.][]{Williams2009}, by following \cite{Straatman2018}. This allows us to more simply distinguish between all galaxies in our sample regardless of redshift \citep[see full details in][]{Straatman2018}. The numbers of star-forming and quenched galaxies classified in this way are shown in Table \ref{table_SAMPLE_env}, along with their numbers split by cosmic web location and also in terms of the group membership of the galaxies (see Section \ref{CW_m_c}). Interestingly, by splitting the sample using $UVJ$ into star-forming and quiescent populations, we find significantly different behaviours in the correlation discussed in \S\ref{mass_environment_}. While for the star-forming population there is only a weak correlation between the stellar mass of galaxies and their environment, for quiescent galaxies the correlation is very strong and thus the correlation found for the full LEGA-C sample as a whole is primarily driven by quiescent galaxies (Figure \ref{Overdensity_distribution}). This likely has physical implications about the tighter link between quiescent galaxies and their environment, which we will explore throughout this paper.

\begin{figure*}
 \centering
 \centering
\begin{tabular}{cc}
\includegraphics[height=7.8cm]{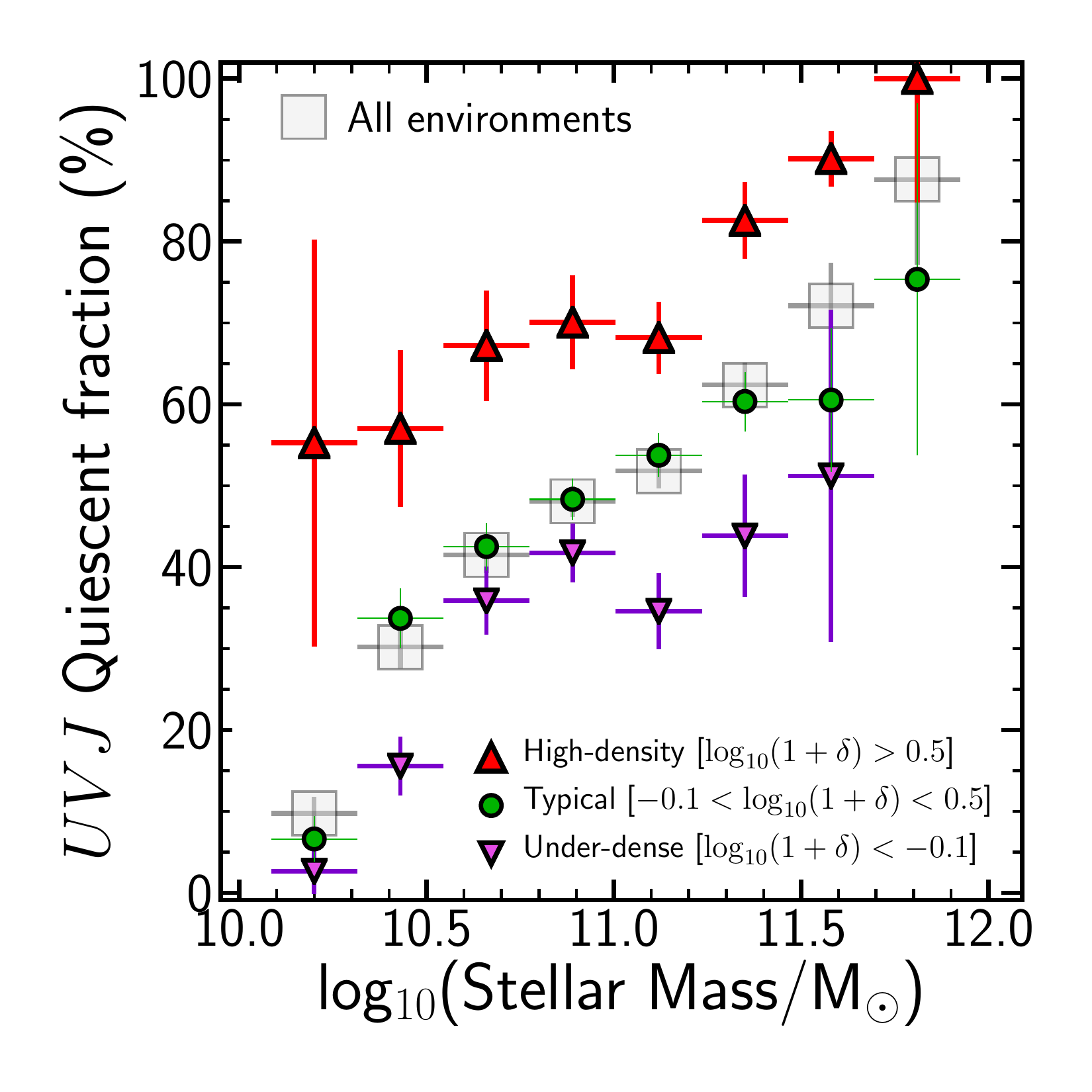}&
\includegraphics[height=7.8cm]{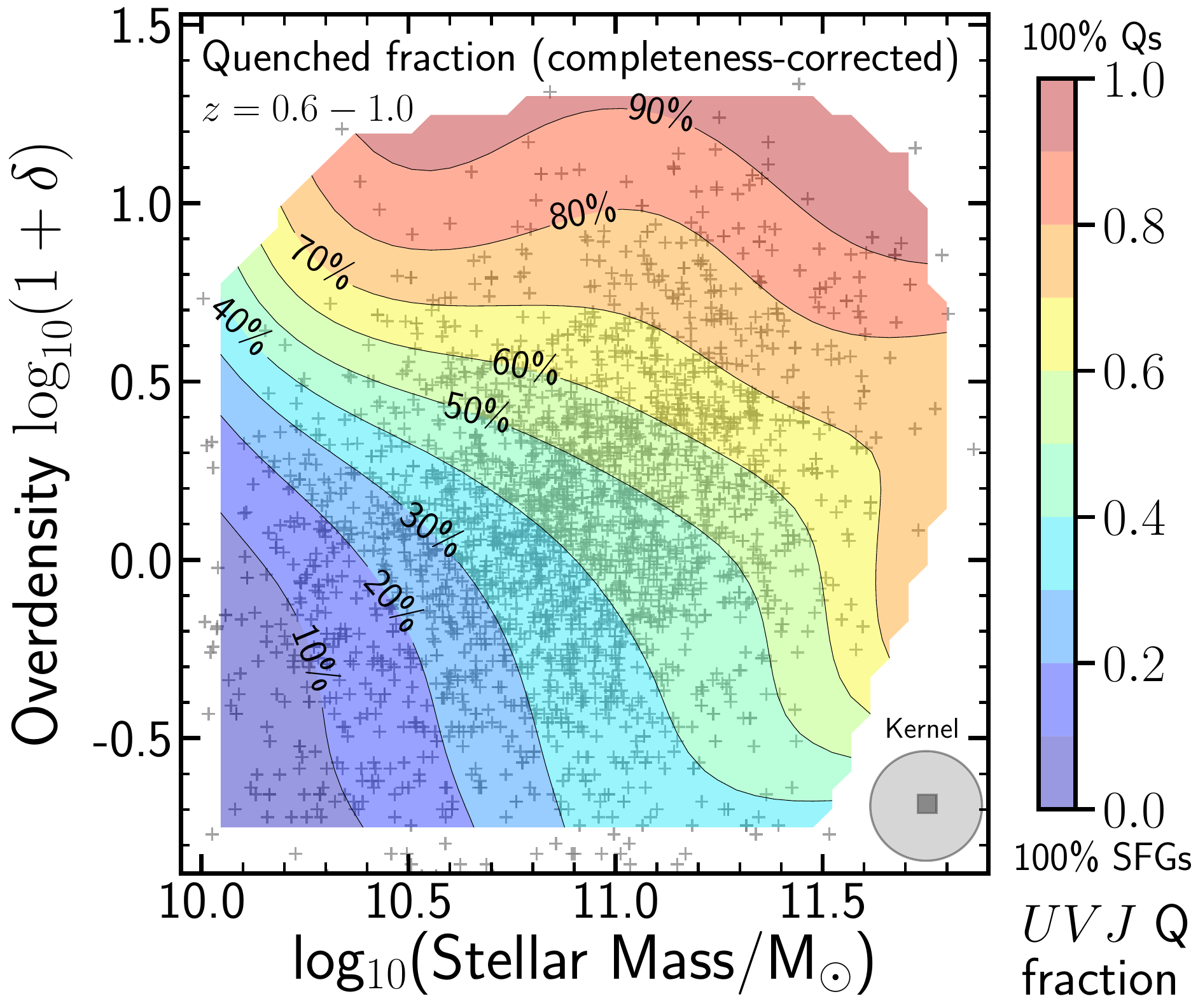}\\
\end{tabular}
 \caption{{\it Left:} The fraction of quiescent galaxies as a function of stellar mass for different environmental densities. Environmental densities were chosen to roughly correspond to the average separation between densities typical of fields, rich fields and filaments, and groups and clusters. We find that the quiescent fraction in LEGA-C at $z\sim0.8$ rises strongly with stellar mass. At fixed stellar mass, the quiescent fraction rises with increasing environmental density, particularly from densities typical of fields/filaments into densities of groups and clusters. Errors are binomial based on the real number of galaxies in LEGA-C (pre-completeness correction). {\it Right:} The fraction of quiescent ($UVJ$ based) galaxies for our LEGA-C sample at $0.6<z<1.0$ as a function of both stellar mass and local environmental density, after correcting for mass completeness. We find strong gradients with both mass and environment, with the quiescent fraction dropping steeply with increasing stellar mass and environment from close to just 10\% to 100\%. We also show the grid and Gaussian kernel used to obtain the quenched fraction in the 2D parameter space, corresponding to $\approx0.2-0.3$ dex in both parameters.}
 \label{Q_fraction_full_sample}
\end{figure*}

In order to estimate fractions of galaxies of different classes (see e.g. Figure \ref{Q_fraction_full_sample}), we apply completeness corrections derived for the LEGA-C survey presented in \cite{Straatman2018}. Fractions are derived as a function of stellar mass for different environments and are related to all galaxies in our sample for that specific stellar mass and environmental bin. We use Poissonian or binomial errors based on the observed numbers.

%%%%%%%%%%%%%%%%%%%%%%%%%%%%%%%%%%%%
%
%  Figure 4 - Fraction of quiescent galaxies as a function of stellar mass, for environmental bins
%  Mass vs Environment: distribution of each and the correlation between both
%
%%%%%%%%%%%%%%%%%%%%%%%%%%%%%%%%%%%%

\subsection{Spectral indices: D$_n$4000 \& H$\delta$}

In this paper we use measurements of D$_n$4000 as a proxy for luminosity-weighted age \citep[see also][]{Balogh1999,Wu2018,Chauke2018} and the strength of H$\delta$ in absorption (in {\AA}) as a proxy for an intense ``recent" burst ($<1$\,Gyr). In this paper, H$\delta$ is measured as the H$\delta_A$ index \citep[e.g.][]{Worthey1997,Balogh1999} and corrected for emission \citep[see e.g.][]{Wu2018}. We will focus on directly observed trends and interpret them in this context. Nevertheless, it is important to note caveats on the potential role of metallicity \citep[see e.g.][]{PA2020} and dust \cite[particularly for star-forming galaxies, see e.g.][]{MacArthur005} in setting D$_n$4000.

Full details of the measurements can be found in \cite{Wu2018} and also in \cite{Bezanson2018}. Briefly, spectral indices are obtained by first modelling the emission lines using the Penalized Pixel-Fitting (pPXF) Python routine \citep{CappellariEmsellem2004,Cappellari2017}. We then subtract the model fits from the observed spectra \citep[see e.g.][]{Bezanson2018} taking into account their velocity dispersions \citep[see also][]{Kuntschner2004}, before measuring indices. To estimate the uncertainties, we take into account both the formal uncertainties based on the noise spectra but we also explore duplicate spectra with $S/N>5$. As reported in \cite{Wu2021}, the uncertainties of D$_n$4000 and H$\delta_A$ from the noise spectra are underestimated by 1.3 and 2.0 times, respectively, and the real typical uncertainties should be $\sim0.03$ and $\sim0.55$\AA, respectively.

\subsection{Quantifying the stellar mass and environmental correlations using partial correlations}

Throughout this work, we quantify the correlation between D$_n$4000 and H$\delta$ on stellar mass and environmental over-density by exploring partial correlations\citep[e.g.][]{Bait2017,Bluck2019}. This method allows us to identify which property (stellar mass and/or environment) is most strongly correlated with D$_n$4000 and H$\delta$ by taking away the effect of the other property. This is particularly important due to stellar mass and environment being intrinsically correlated (see \S\ref{mass_environment_}). In order to make sure that this specific analysis is as unbiased as possible, we obtain partial correlations with sub-samples of stellar mass of $>10^{10.5}$\,M$_{\odot}$. We will report on the correlation strength ($r$), its 95\% confidence interval (C.I.) and identify a partial correlation as significant when $r$ is clearly above 0 within the 95\% confidence interval; see Tables \ref{table_D4000_Mass_environment} and \ref{table_Hdelta_Mass_environment}. In order to visualise the results, we will also show them as arrows which point towards the direction in the Stellar Mass-Environment space in which D$_n$4000 and H$\delta$ are found to change; grey arrows allow us to visualise the 95\% confidence intervals (see e.g. Figure \ref{D4000_Hd_vs_mass_Env}). While the partial correlation coefficients do not capture the full 2-d information of the stellar-mass vs environment plane, they enable us to assess the statistical significance of the global trends.

\section{Results}\label{results}

\subsection{The dependence of the quenched fraction on the environment and stellar mass}\label{Q_fract_all}

Figure \ref{Q_fraction_full_sample} presents the fraction of quiescent galaxies (see Section \ref{Sample} for completeness corrections applied) as a function of mass for different global environments and for the full sample. We find that the $UVJ$-based quiescent/quenched fraction rises steeply with mass, in agreement with previous studies \citep[e.g.][]{,Muzzin12,vanDerBurg2013}. Furthermore, we find that the normalisation of the relation depends significantly on environment: at the lowest environmental densities the quiescent fraction is systematically low at fixed mass and systematically high for high densities \citep[][]{Sobral11,vanDerBurg2013}. The inverse is also true: at fixed environment the quiescent fraction is higher for more massive galaxies. Mass quenching is therefore more efficient in denser environments and the environment more efficiently quench massive galaxies\footnote{Massive galaxies are typically older and hence have had longer to quench.} than low-mass systems \citep[][]{Darvish2016}.

This is consistent with processes that have accelerated galaxy evolution in progressively higher density environments. Furthermore, such trends can also be caused by a higher dominance of satellite galaxies in higher densities, thus rising the overall normalisation of the quiescent fraction.

Given the wide range of stellar masses and environments probed by our LEGA-C sample at $z\sim0.6-1.0$, we can also calculate the quiescent fraction in a more continuous/smooth way when varying either or both mass and environmental density. Figure \ref{Q_fraction_full_sample} (right panel) presents the results, with the quiescent or quenched fraction ($UVJ$-based) as a function of both over-density and stellar mass for our LEGA-C sample at $z\sim0.6-1.0$. We find that, just like for the local Universe \citep[e.g.][]{Peng10}, the quenched fraction is a strong function of both stellar mass and environment, in qualitative agreement with the H$\alpha$-based study at $z\sim0.8$ by \cite{Sobral11}. We also check that our completeness corrections can recover the results from the more general COSMOS population at the mass and environmental density ranges that we are focusing on and thus our results should be representative for those.

The results presented in Figure \ref{Q_fraction_full_sample} could already be obtained with photometric samples and they are meant to provide a baseline for the spectroscopic measurements that, with the large statistics and S/N, are only now possible, thanks to LEGA-C. Here we will focus on the high S/N D$_n$4000 and H$\delta$ measurements for a large and representative population of massive galaxies when the Universe was about half its age, which will allow us to start exploring potential differences in their star formation histories., even if they are small.

%%%%%%%%%%%%%%%%%%%%%%%%%%%%%%%%%%%%
%
%  Figure 5 -  Mass vs Environment and D4000 (left) and Hdelta (right): past and recent SF
%
%%%%%%%%%%%%%%%%%%%%%%%%%%%%%%%%%%%%
\begin{figure*}
 \centering
 \centering
\begin{tabular}{cc}
\includegraphics[width=8.5cm]{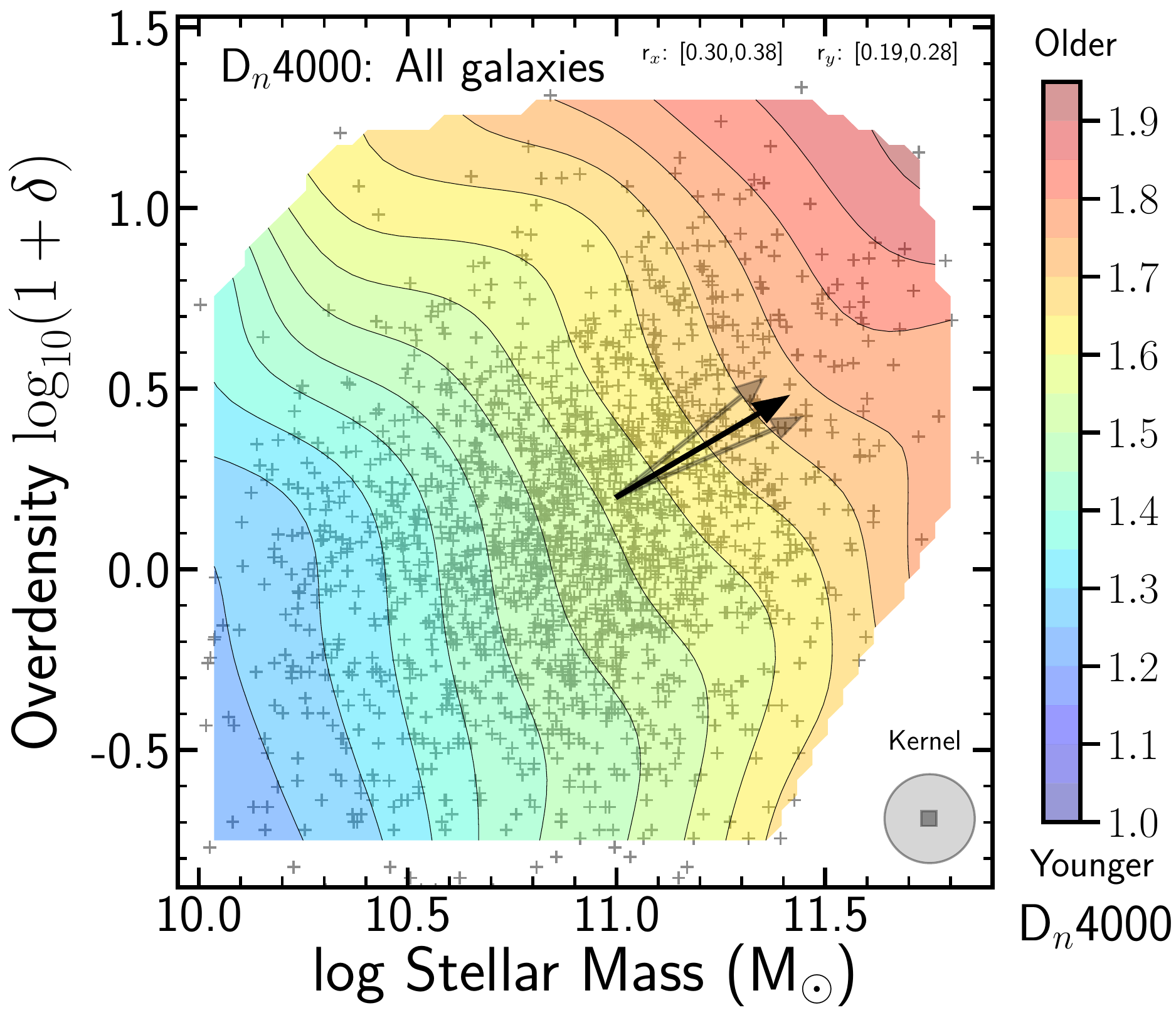}&
\includegraphics[width=8.5cm]{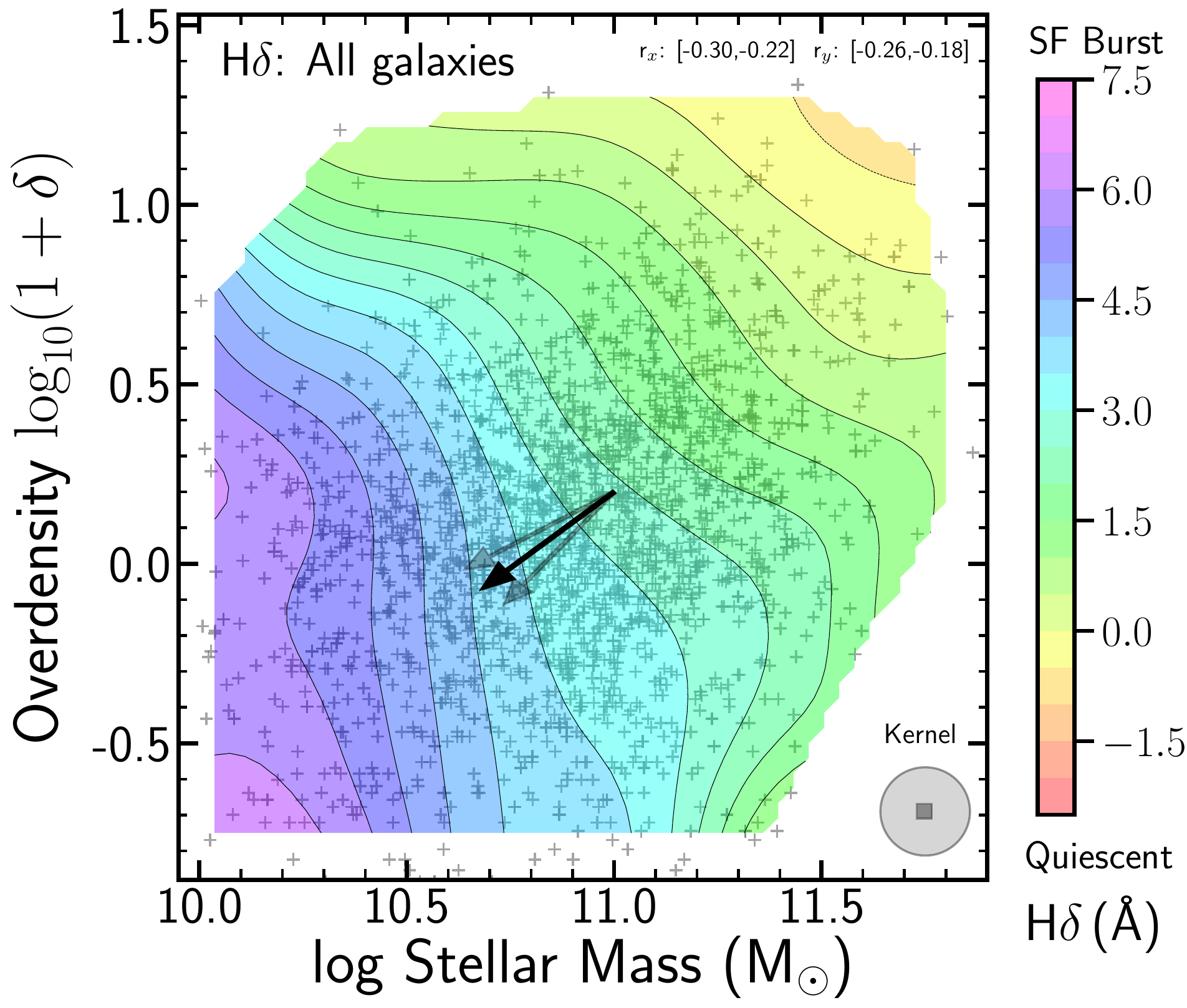}\\
\end{tabular}
 \caption{The D$_n$4000 ({\it left}) and H$\delta$ EW ({\it right}) indices as a function of stellar mass and environmental over-density for our full LEGA-C sample at $z=0.6-1.0$. Both D$_n$4000 and H$\delta$ correlate the strongest with stellar mass for any environment, but both D$_n$4000 and H$\delta$ have significant environmental dependencies at fixed mass, particularly towards rich groups/clusters. This suggests that for a fixed mass, galaxies residing in over-densities formed their stars at earlier times and/or quenched earlier. Results from the partial correlation fitting (see Tables \ref{table_D4000_Mass_environment} and \ref{table_Hdelta_Mass_environment}) are shown with the arrows, where the black arrow shows the combined correlation and the grey arrows the 95\% confidence interval.}
 \label{D4000_Hd_vs_mass_Env}
\end{figure*}

%%%%%%%%%%%%%%%%%%%% TABLE 3   %%%%%%%%%%%%%%%%%%%%%%
%
%         Fitting relations as a function of both mass and environment
%
%%%%%%%%%%%%%%%%%%%%%%%%%%%%%%%%%%%%%%%%%%%%%%%%%
\begin{table*}
\centering
\caption{Partial correlation results when fitting D$_n$4000 as a function of both stellar mass and environment. Note that for all partial correlation analysis we restrict the sample to galaxies with $>10^{10.5}$M$_{\odot}$ where LEGA-C is complete irrespective of the sub-sample. Our analysis allows us to describe the relationship between two variables whilst taking away the effects of another variable: $E$: By removing any environmental over-density dependency and $M$ by removing any stellar mass correlation.}
\begin{tabular}{cccccc}
\hline
D$_{4000}$ & \# of sources   &  D$_{4000}$-Stellar Mass$^E$  & D$_{4000}$-$\log_{10}(1+\delta)^{M}$  & Mass/Environment & Statistical correlation  \\ 
Sub-sample  & \#   &  $r$ (95\% C.I.)  &  $r$ (95\% C.I.)  & (ratio of $r$)  &   \\  \hline

All &  1648  &  $0.34^{+0.04}_{-0.04}$ \checkmark & $0.23^{+0.05}_{-0.04}$  \checkmark  &  1.5 & Mass and Environment \\ 
Centrals &  1129  &  $0.31^{+0.05}_{-0.05}$\checkmark  & $0.25^{+0.06}_{-0.05}$\checkmark    &  1.2 & Mass and Environment \\ 
Satellites &  519  &  $0.27^{+0.08}_{-0.08}$\checkmark  & $0.29^{+0.07}_{-0.09}$\checkmark    &  1.0 & Mass and Environment \\ 
 \hline
 
SFGs &  827  &  $0.43^{+0.06}_{-0.06}$  \checkmark & $0.05^{+0.07}_{-0.07}$ $\times$  &  $>4$ & Mass only \\ 
SFGs-centrals &  555  &  $0.47^{+0.06}_{-0.07}$  \checkmark & $0.08^{+0.08}_{-0.09}$ $\times$   &  $>4$ & Mass only \\ 
SFGs-satellites &  272  &  $0.30^{+0.11}_{-0.11}$  \checkmark & $0.0^{+0.1}_{-0.1}$ $\times$   &  $>4$ & Mass only \\ 
 \hline
Quiescent &  821  &  $0.26^{+0.06}_{-0.06}$  \checkmark & $0.19^{+0.06}_{-0.07}$ \checkmark   &  1.4 & Mass and Environment \\ 
Q-centrals &  574  &  $0.24^{+0.07}_{-0.08}$  \checkmark & $0.16^{+0.08}_{-0.08}$ \checkmark   &  1.5 & Mass and Environment \\ 
Q-satellites &  247  &  $0.23^{+0.12}_{-0.12}$  \checkmark & $0.28^{+0.11}_{-0.12}$ \checkmark   &  0.8 & Mass and Environment \\ 
\hline
\label{table_D4000_Mass_environment}
\end{tabular}
\end{table*}

\subsection{The dependence of D$_n$4000 and H$\delta$ on environment and stellar mass}

Figure \ref{D4000_Hd_vs_mass_Env} shows how D$_n$4000 and H$\delta$ depend on stellar mass and environmental over-density at $z=0.6-1.0$ with LEGA-C. The clear trends with mass at $z\sim0.6-1.0$ have already been explored by \cite{Wu2018}, who found that D$_n$4000 rises with increasing stellar mass, while H$\delta$ decreases with increasing mass. Here we show that both D$_n$4000 and H$\delta$ depend on environment at fixed mass, even though the correlation with environment is not as strong as with stellar mass (see Tables \ref{table_D4000_Mass_environment} and \ref{table_Hdelta_Mass_environment} for the quantitative results).

At the lower stellar masses probed, galaxies have the highest H$\delta$ EWs and the lowest D$_n$4000, implying young ages and recent episodes of star-formation. Below $10^{10.5}$\,M$_{\odot}$ we still find tentative environmental trends of increasing D$_n$4000 and decreasing H$\delta$ EW at fixed mass, but the trends are much stronger and statistically much more significant for higher mass galaxies. The dependence on environment is also seen to be stronger above environmental densities linked with rich fields into groups and clusters, while at typical densities (and under-densities), the gradient with environment (a change along the y axis in Figure \ref{D4000_Hd_vs_mass_Env}) is very weak to non-existent. 

%%%%%%%%%%%%%%%%%%%%%%%%%%%%%%%%%%%%
%
%  Figure 6 -  Mass vs Environment and D4000 (left) and Hdelta (right): past and recent SF
%
%%%%%%%%%%%%%%%%%%%%%%%%%%%%%%%%%%%%
\begin{figure*}
 \centering
 \centering
\begin{tabular}{cc}
\includegraphics[scale=0.43]{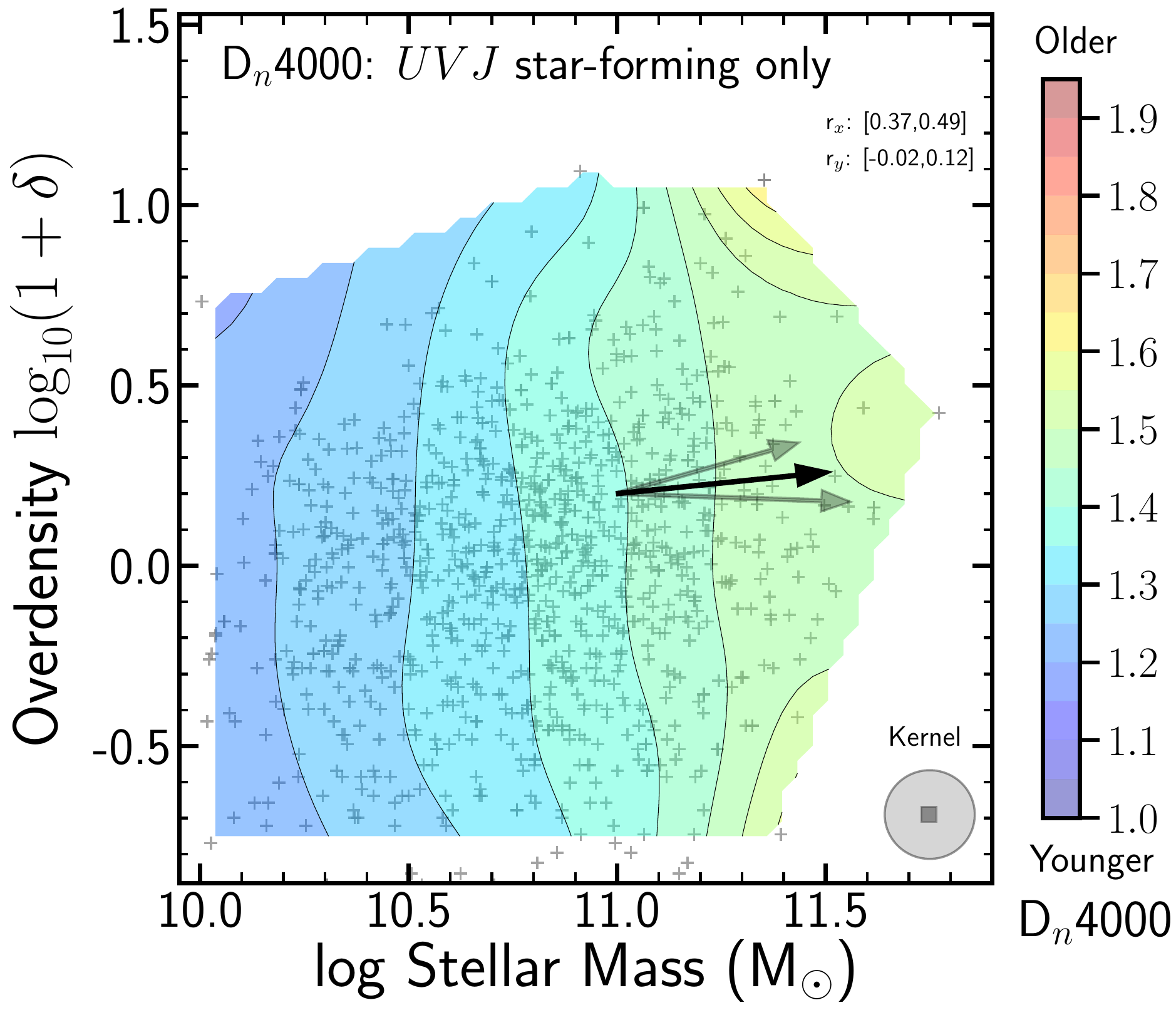}&
\includegraphics[scale=0.43]{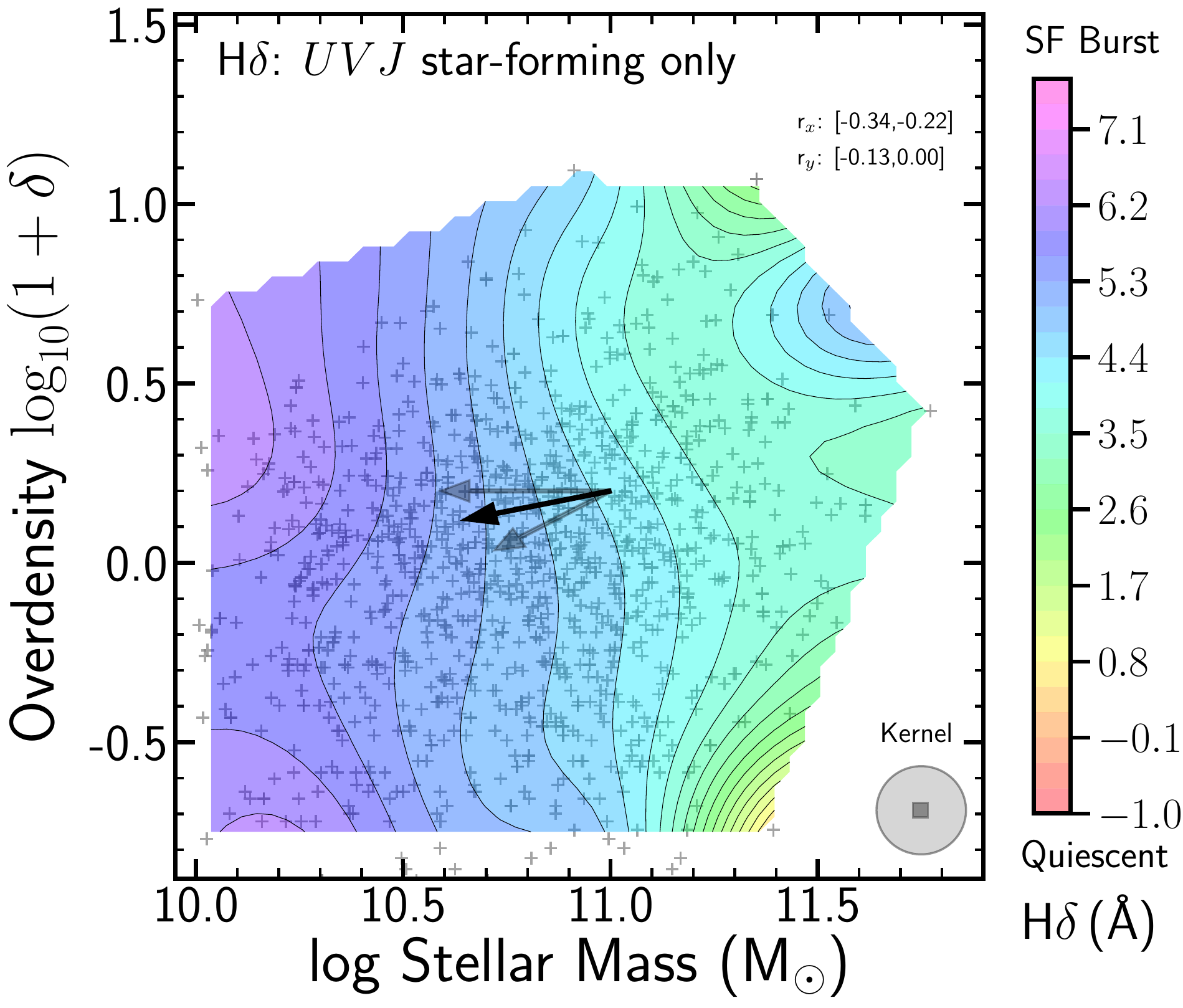}\\
\end{tabular}
\begin{tabular}{cc}
\includegraphics[scale=0.43]{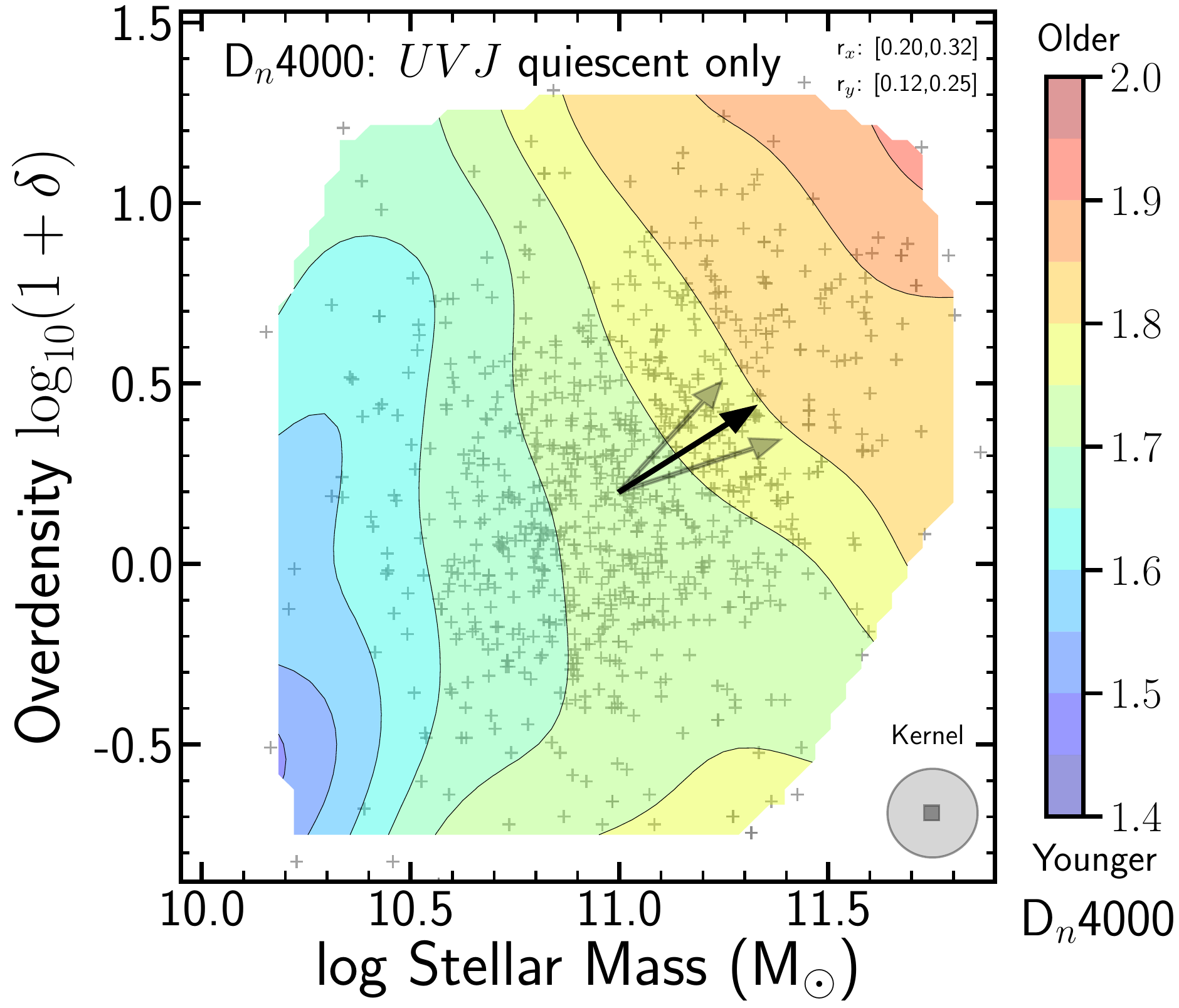}&
\includegraphics[scale=0.43]{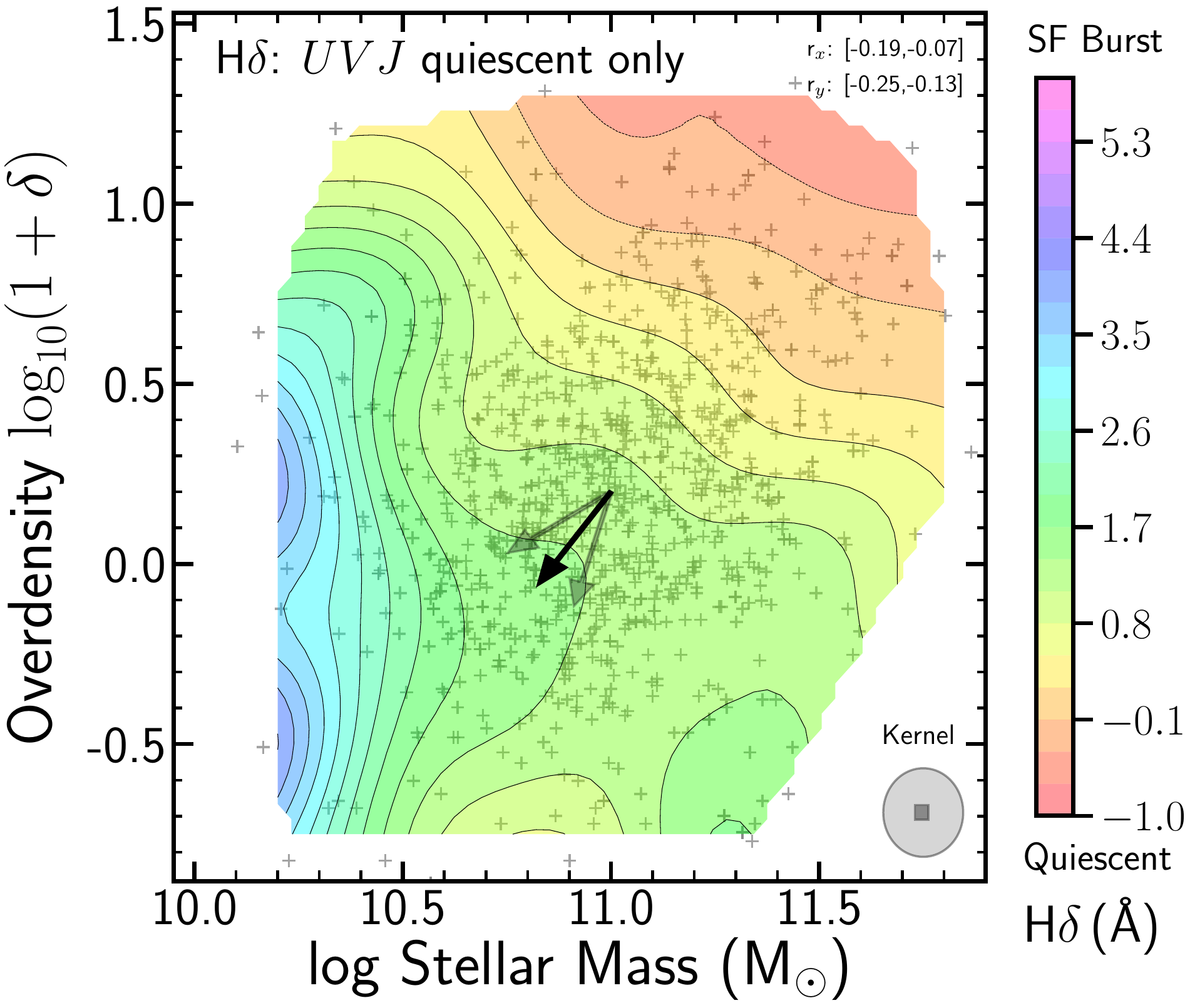}\\
\end{tabular}
 \caption{The average D$_n$4000 ({\it left}) and H$\delta$ ({\it right}) as a function of stellar mass and environmental over-density for star-forming galaxies only (top) and for quiescent galaxies only (bottom) at $z=0.6-1.0$. For star-forming galaxies, D$_n$4000 and H$\delta$ show no trend with environment, while for quiescent galaxies, both indices show important environmental trends.}
 \label{D4000_Hd_vs_mass_Env_SF_Q}
\end{figure*}

Galaxies with the highest stellar masses of $>10^{11}$\,M$_{\odot}$ have the lowest H$\delta$ EWs and the highest D$_n$4000, implying that they are the oldest and least star-forming (relative to their stellar mass), which has been well established. However, our results reveal that at fixed mass, high mass galaxies residing in higher density environments such as groups and clusters have higher D$_n$4000 and even lower H$\delta$ EWs than their counterparts residing in typical density regions. This means that we are likely seeing evidence for different assembly histories, with galaxies at $z\sim0.6-1.0$ having potentially formed earlier and/or faster in high-density regions than galaxies with similar mass in lower-density environments.

%%%%%%%%%%%%%%%%%%%%%%%%%%%%%%%%%%%%
%
%  Figure 7 -     First plot with D4000 and Hdelta centrals top, satellites below
%
%%%%%%%%%%%%%%%%%%%%%%%%%%%%%%%%%%%%
\begin{figure*}
 \centering
 \centering
\begin{tabular}{cc}
\includegraphics[scale=0.42]{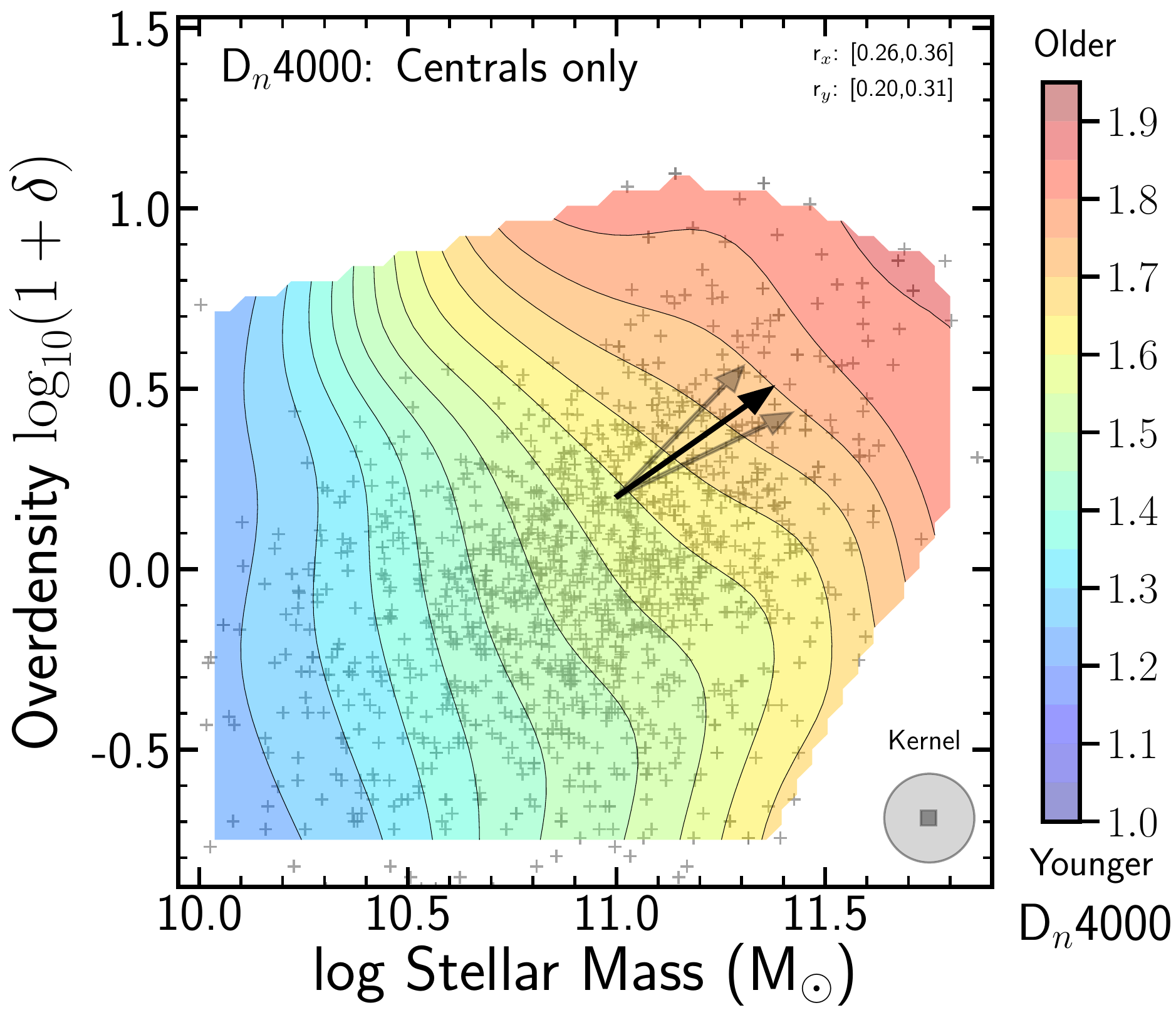}&
\includegraphics[scale=0.42]{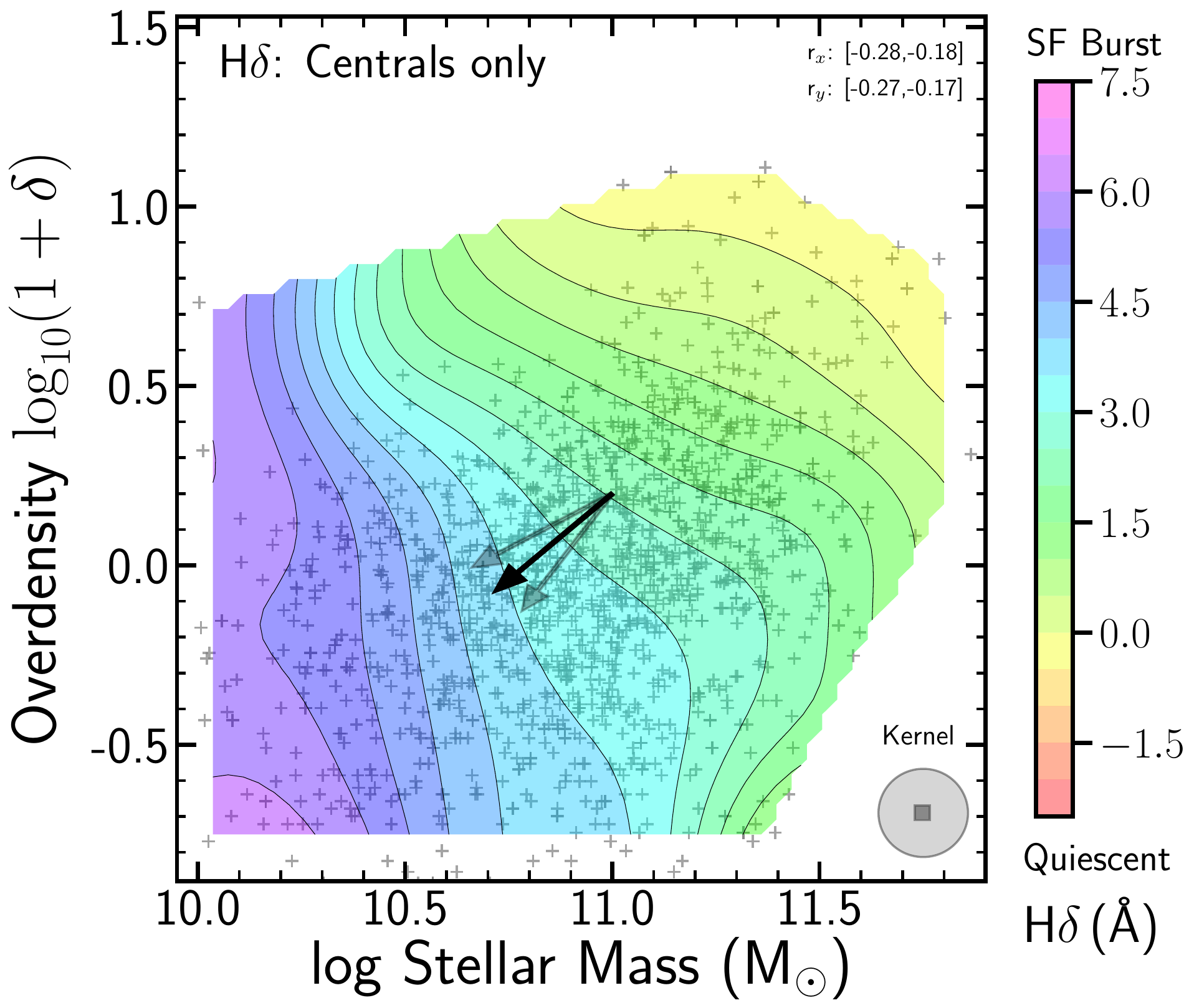}\\
\end{tabular}
\begin{tabular}{cc}
\includegraphics[scale=0.425]{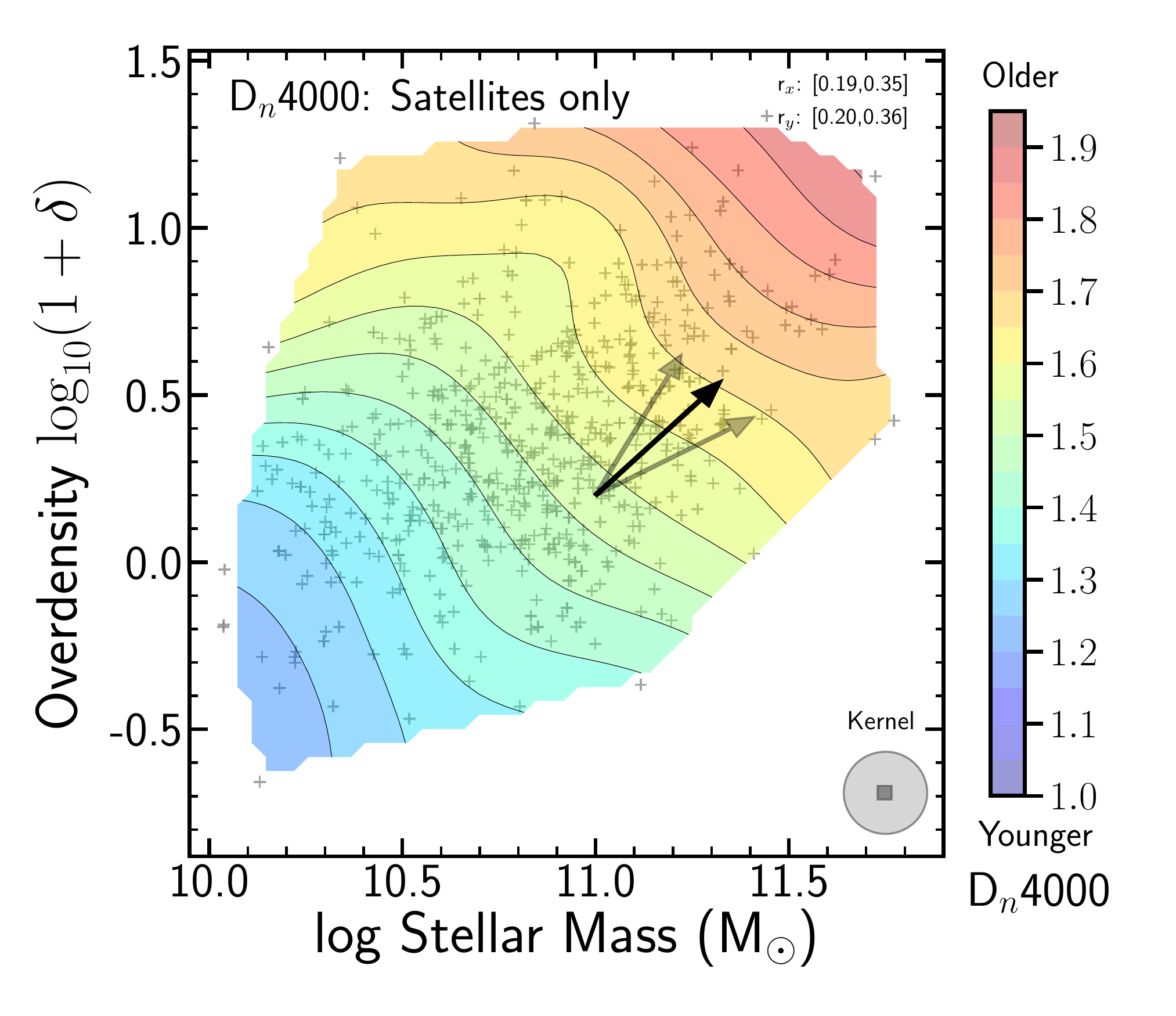}&
\includegraphics[scale=0.42]{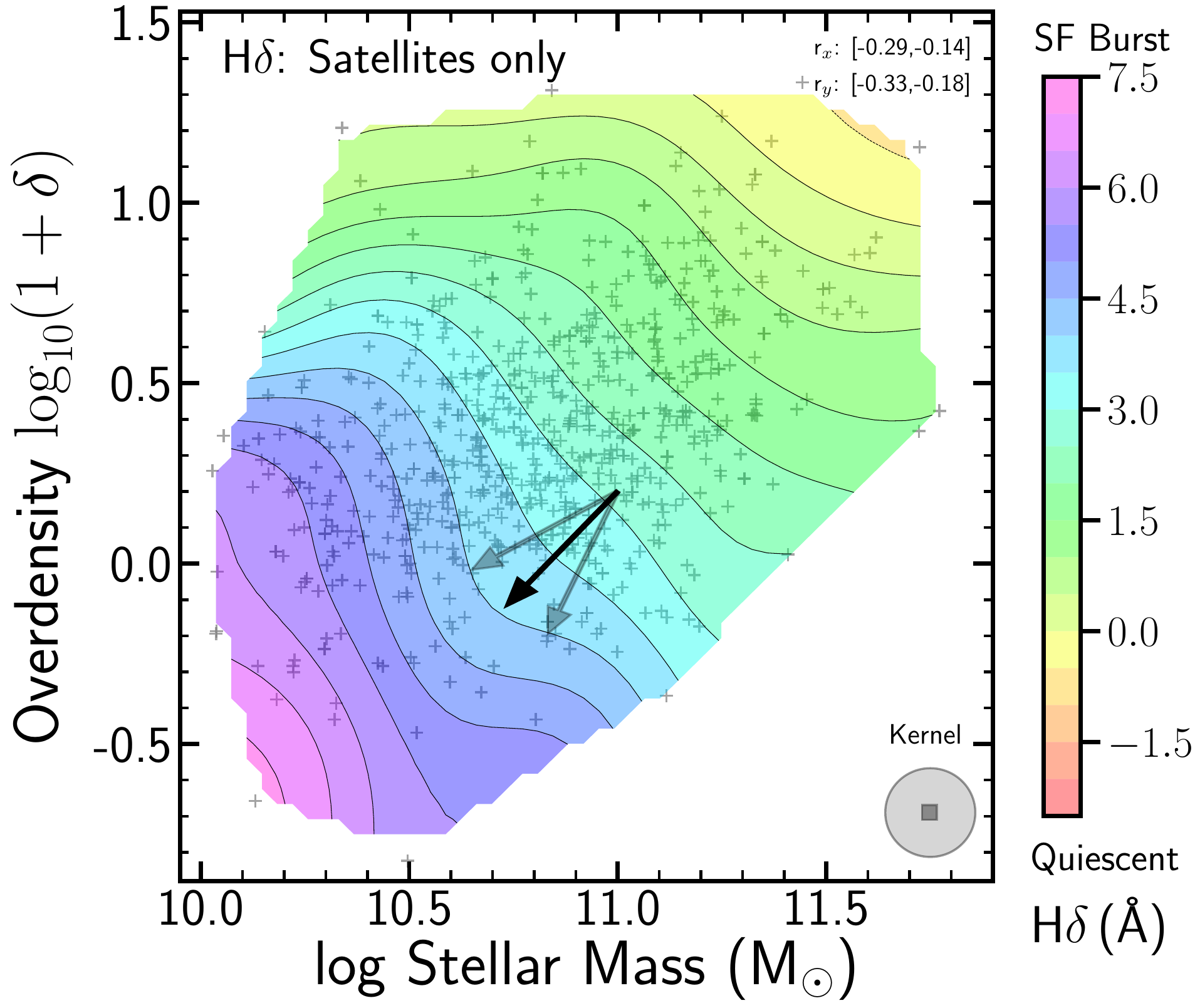}\\
\end{tabular}
 \caption{The average D$_n$4000 ({\it left}) and H$\delta$ ({\it right}) as a function of stellar mass and environmental over-density for centrals only (top) and for satellites (bottom). Both D$_n$4000 and H$\delta$ correlate the strongest with stellar mass for any environment. However, D$_n$4000 and H$\delta$ have significant environmental dependencies at fixed mass, which are stronger for satellites, but are still seen for high mass centrals.}
 \label{D4000_Hd_vs_mass_Env_Centrals_vs_SATS}
\end{figure*}

%%%%%%%%%%%%%%%%%%%% TABLE 4   %%%%%%%%%%%%%%%%%%%%%%
%
%         Fitting relations as a function of both mass and environment
%
%%%%%%%%%%%%%%%%%%%%%%%%%%%%%%%%%%%%%%%%%%%%%%%%%
\begin{table*}
\centering
\caption{Partial correlation results when fitting H$\delta$ as a function of both stellar mass and environment. Note that for all partial correlation analysis we restrict the sample to galaxies with $>10^{10.5}$M$_{\odot}$ where LEGA-C is complete irrespective of the sub-sample. Our analysis allows us to describe the relationship between two variables whilst taking away the effects of another variable: $E$: By removing any environmental over-density dependency and $M$ by removing any stellar mass correlation.}
\begin{tabular}{cccccc}
\hline
H$\delta$ & \# of sources   &  H$\delta$-Stellar Mass$^E$  & H$\delta$-$\log_{10}(1+\delta)^M$  & Mass/Environment & Statistical correlation  \\ 
Sub-sample  & \#   &  $r$ (95\% C.I.)  &  $r$ (95\% C.I.)  & (ratio of $r$)  &   \\  \hline
All &  1851  &  $-0.26^{+0.04}_{-0.04}$  \checkmark & $-0.22^{+0.04}_{-0.04}$  \checkmark   &  1.2  & Mass and Environment \\ 
Centrals &  1262  &  $-0.23^{+0.05}_{-0.05}$   \checkmark & $-0.22^{+0.05}_{-0.05}$   \checkmark   &  1.0   & Mass and Environment \\ 
Satellites &  589  &  $-0.22^{+0.08}_{-0.07}$   \checkmark & $-0.26^{+0.08}_{-0.07}$   \checkmark   &  0.8    & Mass and Environment \\ 
\hline
SFGs &  920  &  $-0.28^{+0.06}_{-0.06}$ \checkmark & $-0.06^{+0.06}_{-0.07}$  $\times$   &  $>3$ & Mass only \\ 
SFGs-centrals &  616  &  $-0.28^{+0.07}_{-0.07}$ \checkmark & $-0.06^{+0.08}_{-0.07}$ $\times$   &  $>3$ & Mass only \\ 
SFGs-satellites &  304  &  $-0.24^{+0.11}_{-0.10}$ \checkmark & $-0.0^{+0.1}_{-0.1}$  $\times$  &  $>3$ & Mass only \\ 
\hline
Quiescent &  931  &  $-0.13^{+0.06}_{-0.06}$ \checkmark & $-0.19^{+0.06}_{-0.06}$ \checkmark   &  0.7 & Mass and Environment \\ 
Q-centrals &  646  &  $-0.11^{+0.08}_{-0.07}$ \checkmark & $-0.16^{+0.08}_{-0.07}$  \checkmark  &  0.7 & Mass and Environment \\ 
Q-satellites &  285  &  $-0.14^{+0.11}_{-0.11}$ \checkmark & $-0.25^{+0.12}_{-0.10}$ \checkmark   &  0.6 & Mass and Environment\\ 
\hline
\label{table_Hdelta_Mass_environment}
\end{tabular}
\end{table*}

%%%%%%%%%%%%%%%%%%%%%%%%%%%%%%%%%%%%
%
%  Figure 8 -     First plot with D4000 and Hdelta centrals top, satellites below, just for Star-forming galaxies
%
%%%%%%%%%%%%%%%%%%%%%%%%%%%%%%%%%%%%
\begin{figure*}
 \centering
 \centering
\begin{tabular}{cc}
\includegraphics[scale=0.43]{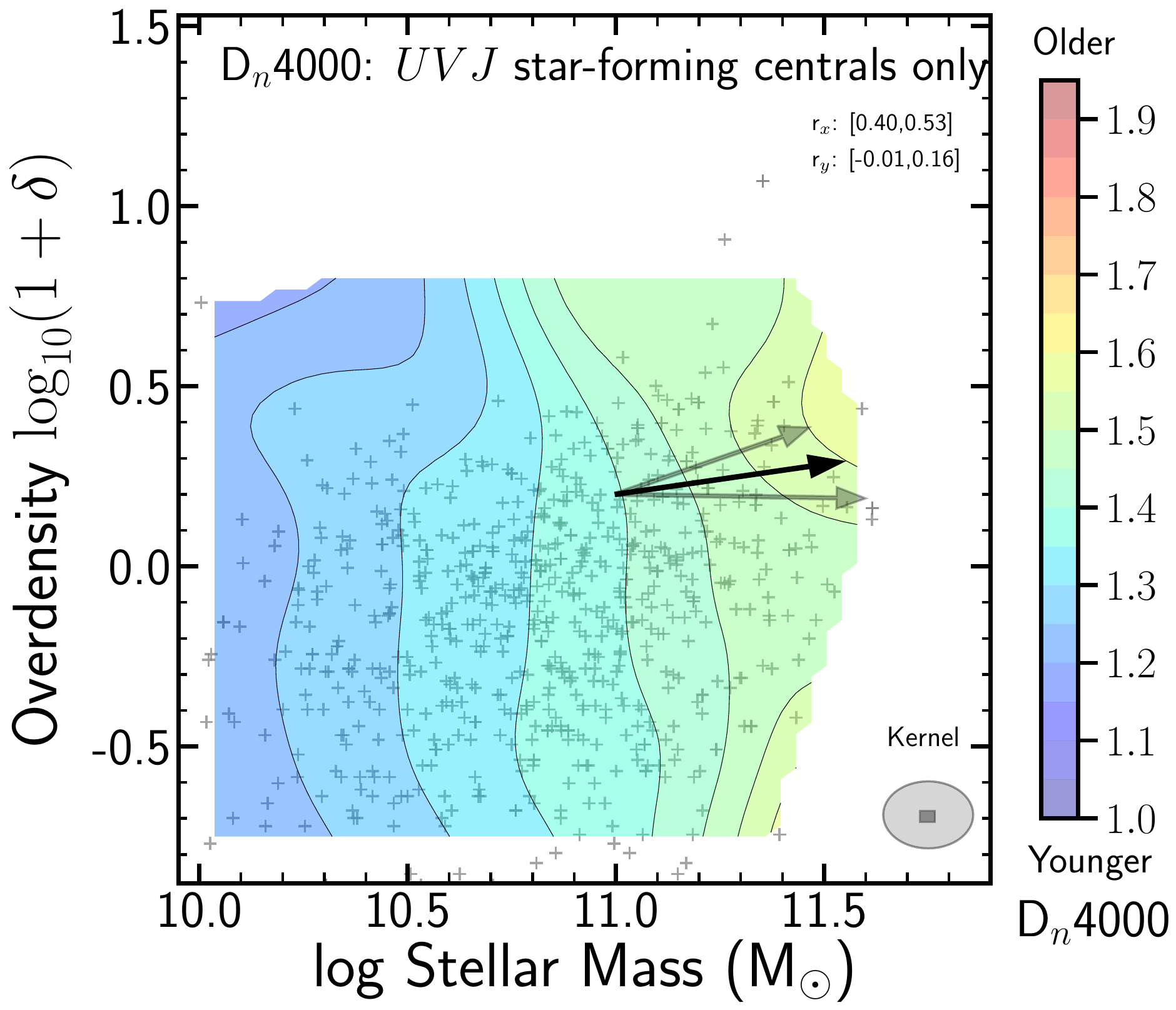}&
\includegraphics[scale=0.43]{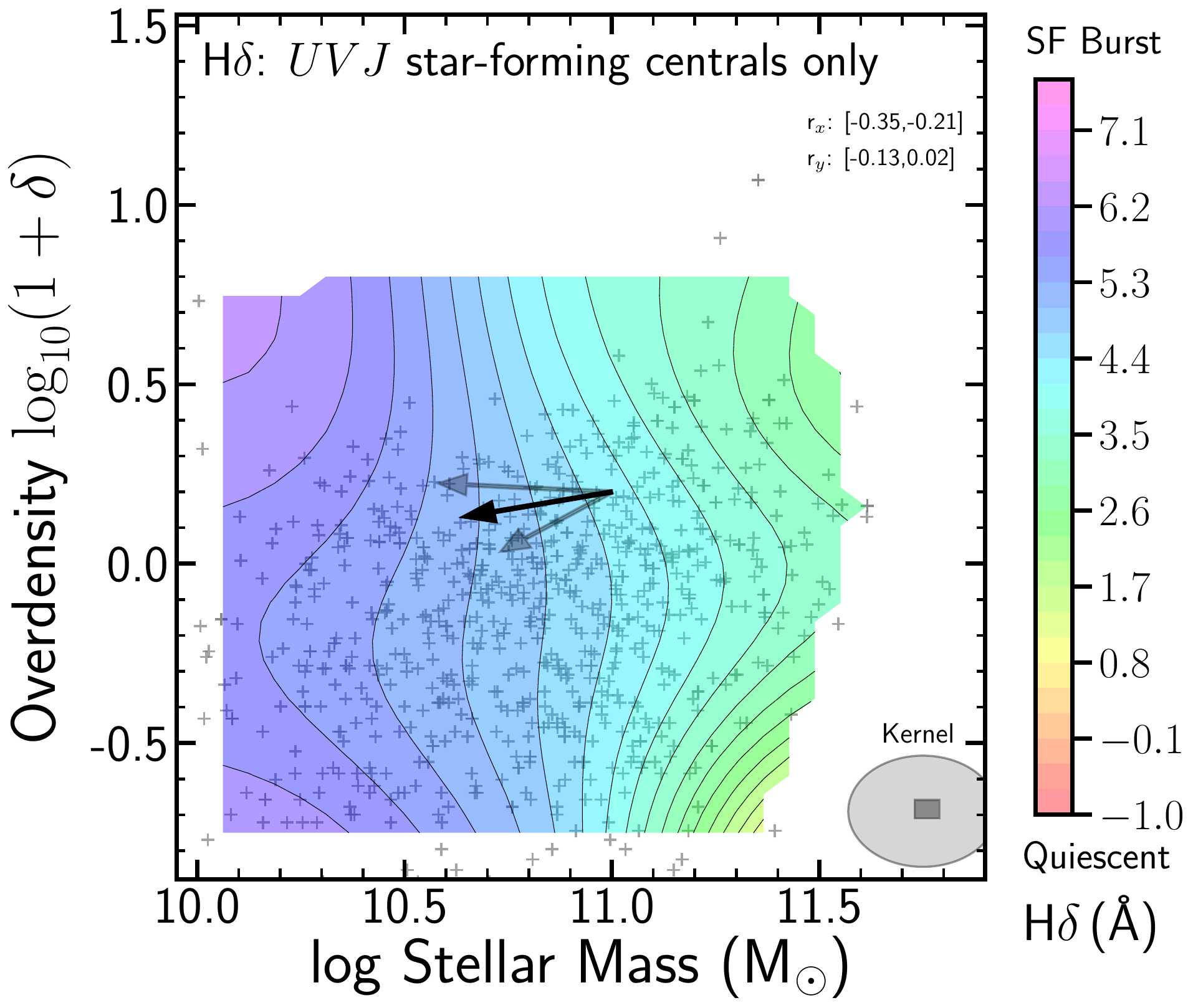}\\
\end{tabular}
\begin{tabular}{cc}
\includegraphics[scale=0.435]{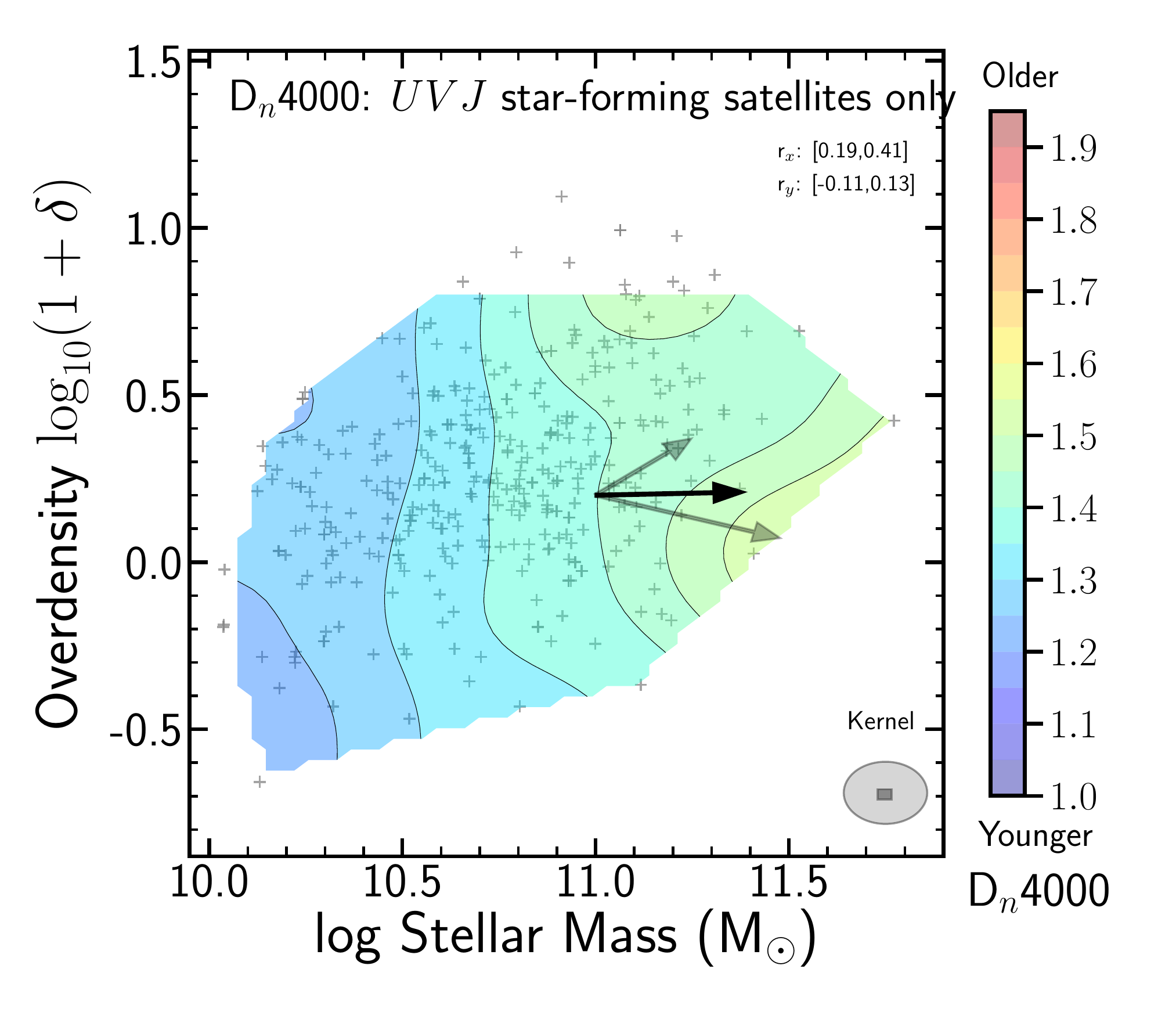}&
\includegraphics[scale=0.43]{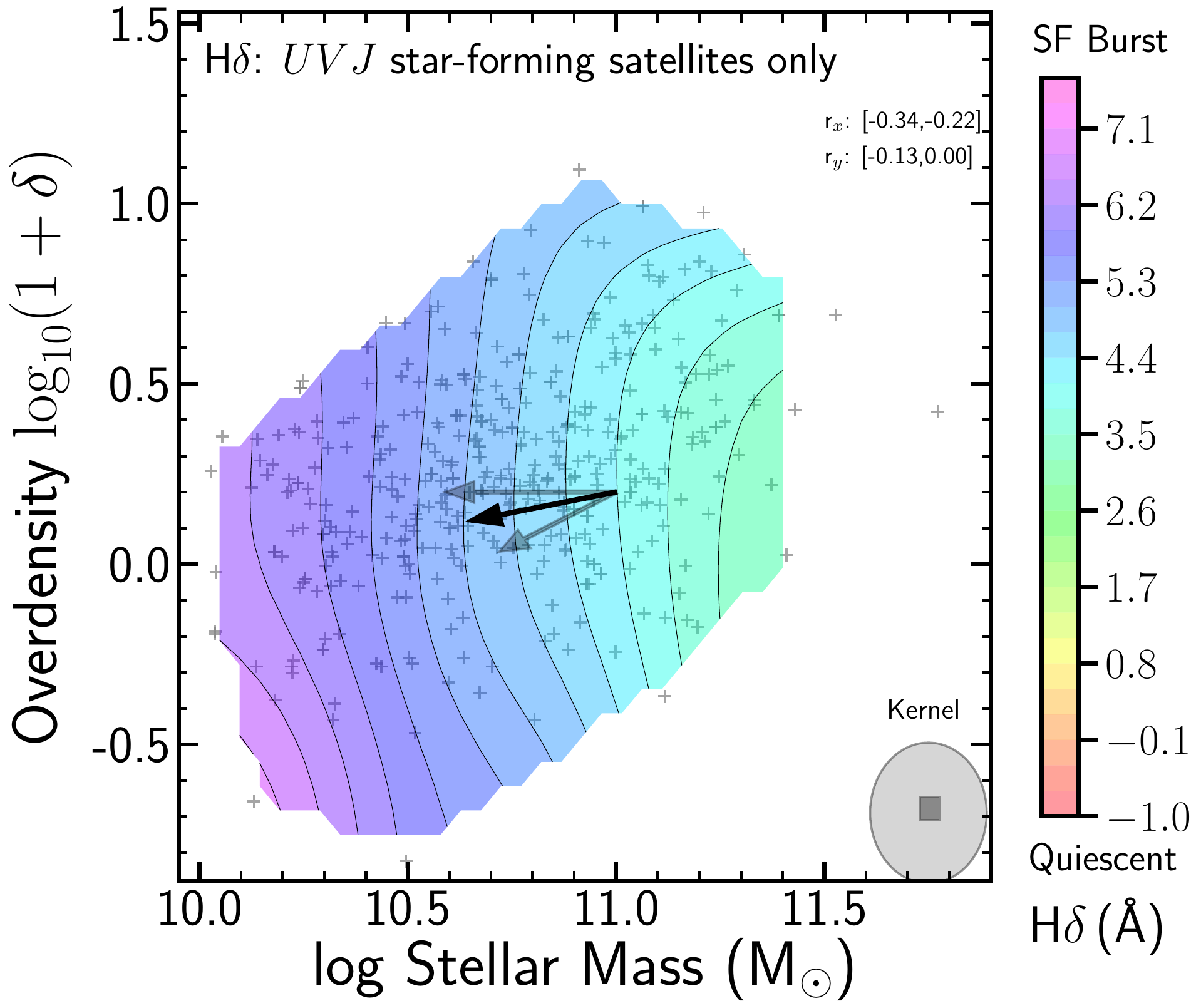}\\
\end{tabular}
 \caption{The average D$_n$4000 ({\it left}) and H$\delta$ ({\it right}) as a function of stellar mass and environmental over-density for star-forming centrals (top) and for star-forming satellites (bottom). Both D$_n$4000 and H$\delta$ correlate with stellar mass, but there are no gradients with environment.}
 \label{D4000_Hd_vs_mass_Env_SFCentrals_SATS}
\end{figure*}

Similar trends can be recovered if we simply split each mass bin into three different broad environments, capturing e.g. i) poor/under-dense fields, ii) fields to filaments and iii) groups, cluster outskirts and cores. At fixed mass, there are systematic differences in the strength of D$_n$4000 and H$\delta$ for different environments. However, these are small in absolute terms, and thus are only revealed in a statistically significant way with a sample such as LEGA-C. For stellar masses of $\approx10^{10.5}$\,M$_{\odot}$ and above, galaxies residing in the highest densities have higher D$_n$4000 and lower H$\delta$ than those residing in lower to under-dense regions, with the offset at fixed mass being typically 0.1\,dex, similar to what has been found for the local Universe \citep[e.g.][]{Thomas2005}. These trends suggest that galaxy formation/evolution may be accelerated as a function of environmental density, leading to median higher ages in higher densities and lower ages for lower densities already 7 Gyrs ago, consistent with predictions from simulations \citep[e.g.][]{Wechsler2018,Martizzi2020} and with observations of the local Universe \citep[][]{Thomas2005}.

\subsection{The dependence of D$_n$4000 and H$\delta$ on environment and stellar mass: SFGs vs Quiescent}

One important aspect to consider is that the strong shift from star-forming dominated to a quiescent-dominated population as a function of both stellar mass and environment (see Figure \ref{Q_fraction_full_sample}) might easily lead to trends in D$_n$4000 and H$\delta$. This might be the case in the simple hypothesis that all star-forming galaxies are very young and have low D$_n$4000 and high H$\delta$ and that quiescent galaxies have high D$_n$4000 and low H$\delta$. The results in Figure \ref{D4000_Hd_vs_mass_Env} could, therefore, result from gradients in the quiescent fraction \citep[e.g.][]{Muzzin12}.

In Figure \ref{D4000_Hd_vs_mass_Env_SF_Q} we split the population in quiescent and star-forming and investigate how D$_n$4000 and H$\delta$ depend on stellar mass and environmental over-density at $z=0.6-1.0$. The results reveal different trends for star-forming and quiescent galaxies.

Star-forming galaxies reveal gradients in D$_n$4000 and H$\delta$ with stellar mass, but no significant trend with environment. This means that massive star-forming galaxies at $z=0.6-1.0$ of fixed mass reveal apparently similar ages and similar burst strengths, whether they are forming stars in under-densities or in over-densities. The gradient in D$_n$4000 with stellar mass for star-forming galaxies is statistically significant, while H$\delta$ also depends strongly on stellar mass. These results are quantitatively conveyed by the partial correlation coefficients in Tables~\ref{table_D4000_Mass_environment} (for D$_n$4000) and  \ref{table_Hdelta_Mass_environment} (for $H\delta$), where we find that controlling for environment, stellar mass gives statistically significant correlations with both indices, but controlling for stellar mass, there is no evidence of correlation between the indices and environment.

In contrast, the quiescent population shows strong D$_n$4000 and H$\delta$ gradients as a function of both stellar mass and environment (Figure \ref{D4000_Hd_vs_mass_Env_SF_Q}). The D$_n$4000 and H$\delta$ values span a much larger range in respect to stellar mass than those for star-forming galaxies. Most interestingly, at fixed stellar mass, massive quiescent galaxies have larger ($\approx+0.1$) D$_n$4000 and lower H$\delta$ EWs ($\approx-2$\,{\AA}) at higher densities when compared to lower densities; see also Tables~\ref{table_D4000_Mass_environment} (for D$_n$4000) and  \ref{table_Hdelta_Mass_environment} (for $H\delta$).

Overall, by splitting the large LEGA-C sample into star-forming and quiescent, we reveal that much of the trends found for the overall population are driven by the trends found for the quiescent population. Quiescent galaxies at fixed mass reveal evidence for being older and formed earlier in over-densities such as clusters and groups when compared to their field counterparts which are on average younger and formed stars more recently. Star-forming galaxies show evidence of being older and less active as a function of mass. However, at fixed mass there are no visible environmental effects for star-forming galaxies (Tables~\ref{table_D4000_Mass_environment} and  \ref{table_Hdelta_Mass_environment}). This could be explained by a potential combination of rapid environmental quenching and/or by the fact that star-forming galaxies have a large enough fraction of new stars to outshine past/older stellar populations and thus hide any signatures of (much) older populations that might have formed early in higher densities than in lower-density regions (see Discussion).

\subsection{D$_n$4000 and H$\delta$: centrals \& satellites}

In order to investigate the potential different environmental effects acting on satellites and central galaxies, we split all sources and investigate any dependency between D$_n$4000 and H$\delta$ with stellar mass and environment. The results are shown in Figure \ref{D4000_Hd_vs_mass_Env_Centrals_vs_SATS}: centrals are shown in the top panels, while satellites are shown in the bottom panels. It is worth re-enforcing that at a look-back time of 7 Gyrs, the difference between satellites and centrals for massive galaxies may not be expected to be striking. This is because satellites have likely only been satellites for significantly less time than satellites of similar mass in the local Universe.

%%%%%%%%%%%%%%%%%%%%%%%%%%%%%%%%%%%%
%
%  Figure 9 -     First plot with D4000 and Hdelta centrals top, satellites below, just for Quiescent galaxies
%
%%%%%%%%%%%%%%%%%%%%%%%%%%%%%%%%%%%%
\begin{figure*}
 \centering
 \centering
\begin{tabular}{cc}
\includegraphics[scale=0.43]{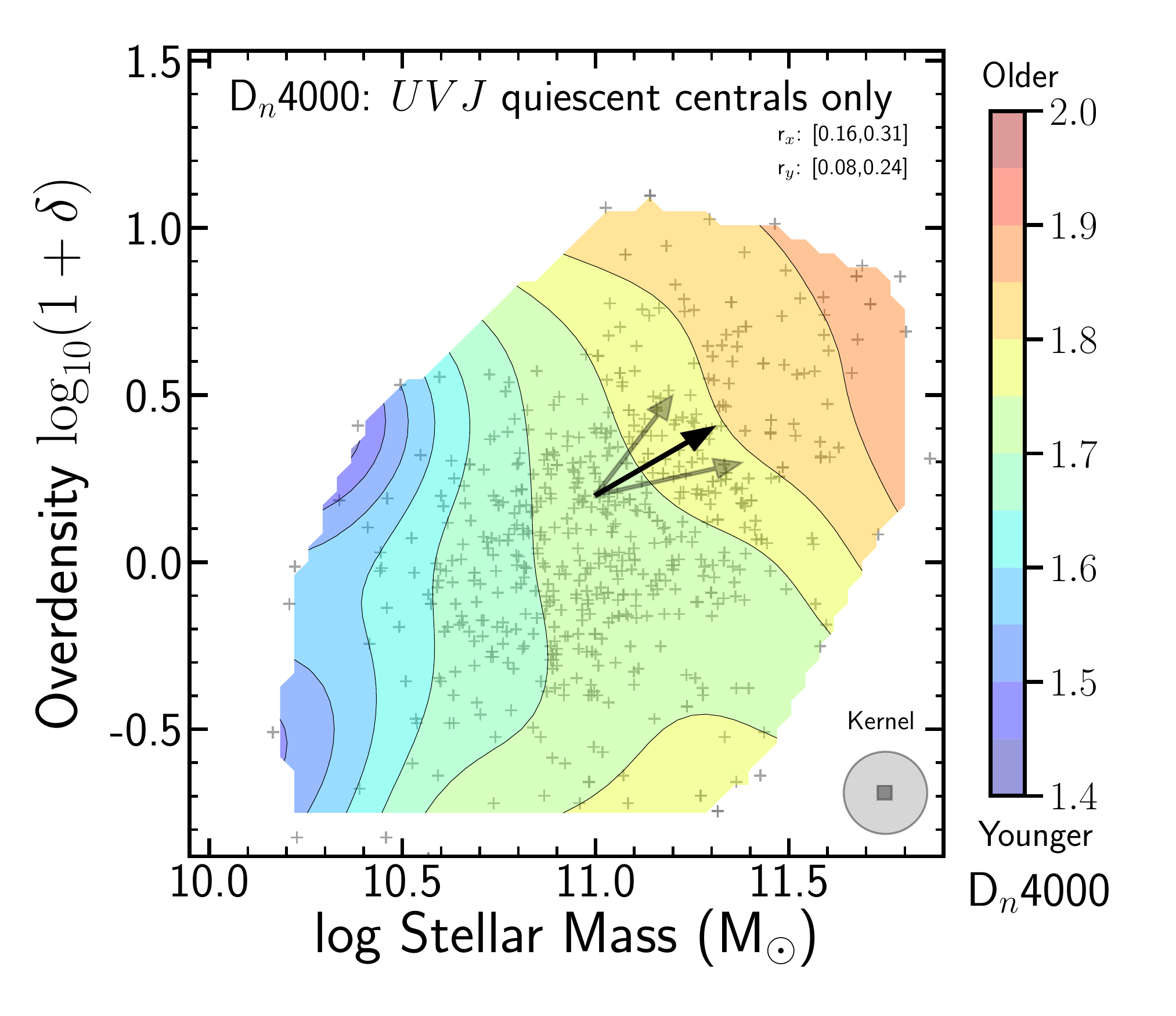}&
\includegraphics[scale=0.43]{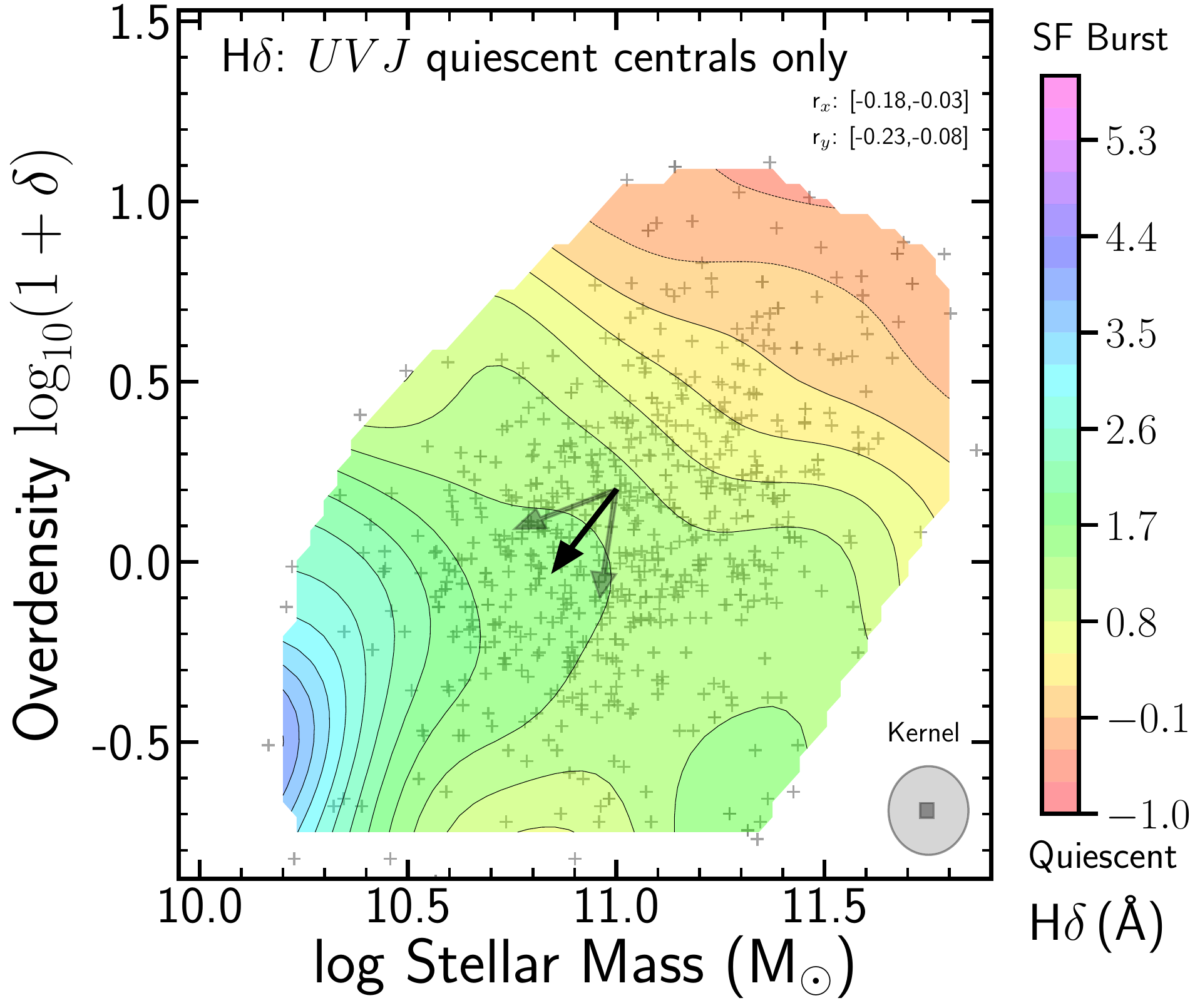}\\
\end{tabular}
\begin{tabular}{cc}
\includegraphics[scale=0.435]{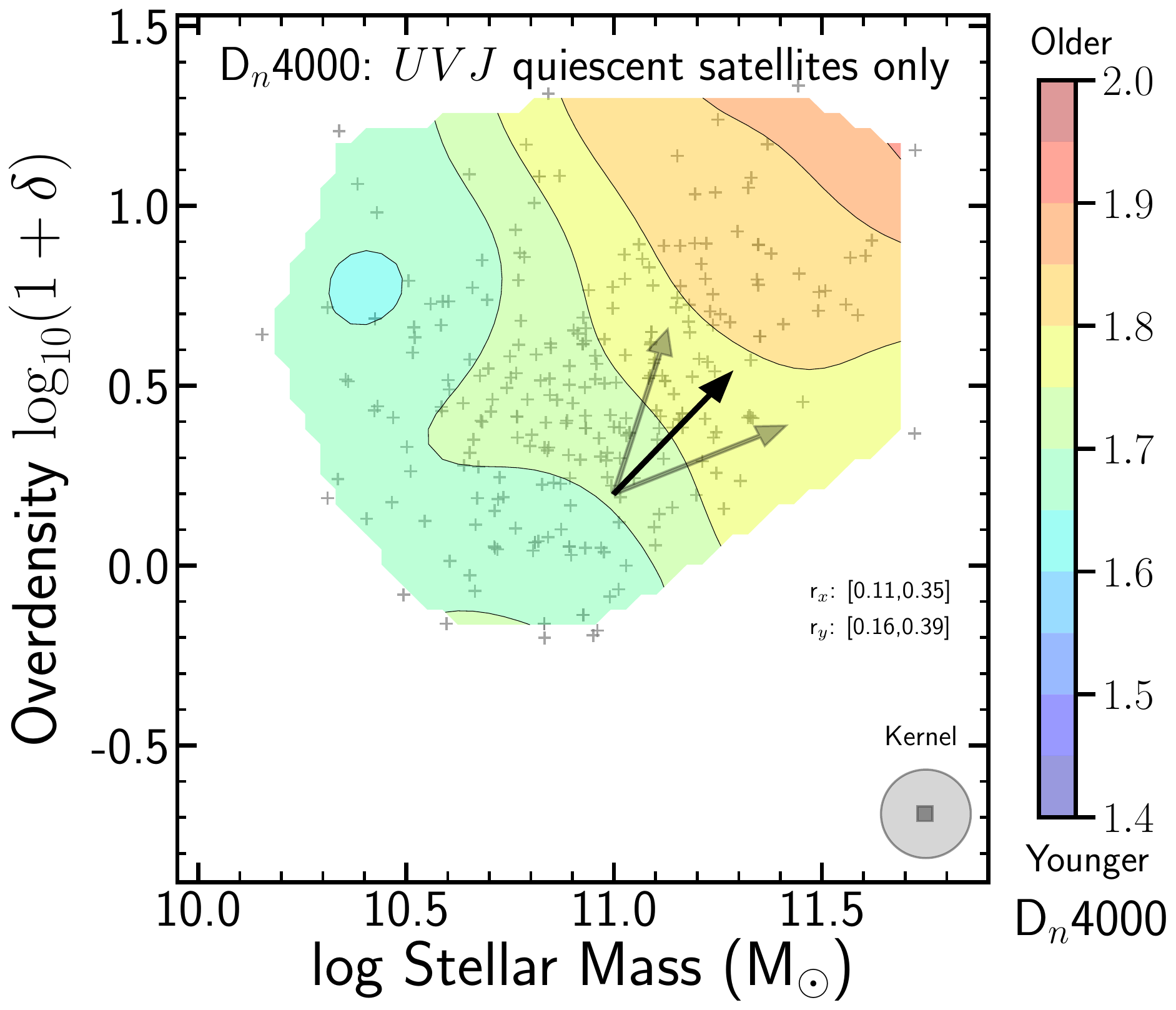}&
\includegraphics[scale=0.43]{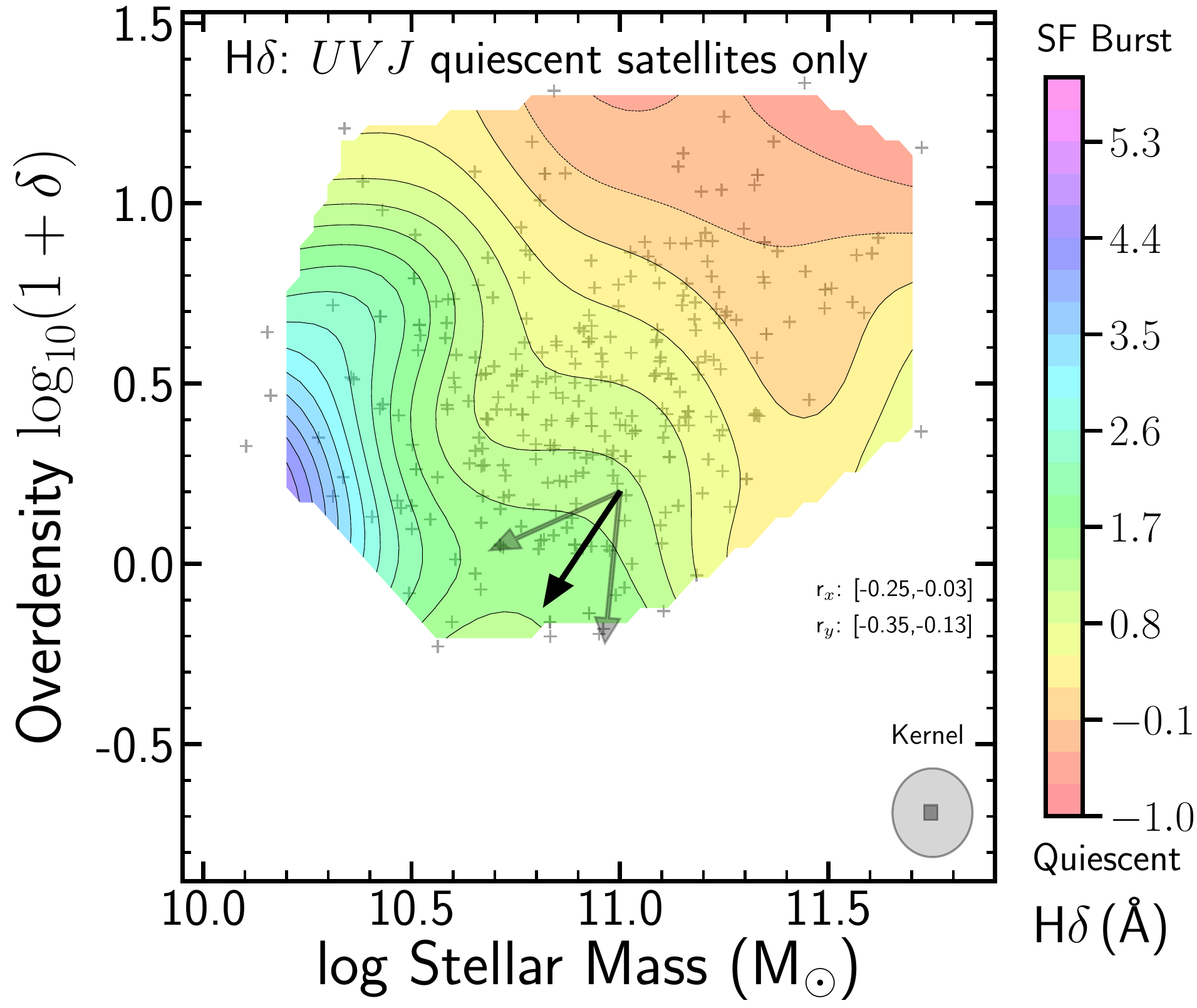}\\
\end{tabular}
 \caption{The average D$_n$4000 ({\it left}) and H$\delta$ ({\it right}) as a function of stellar mass and environmental over-density for quiescent centrals (top) and quiescent satellites (bottom). Both D$_n$4000 and H$\delta$ correlate the strongest with stellar mass for any environment, but both D$_n$4000 and H$\delta$ have significant environmental dependencies at fixed mass. The stellar mass of quiescent centrals is highly correlated with the environment it resides in.}
 \label{D4000_Hd_vs_mass_Env_Qcentrals_vs_Sats}
\end{figure*}

Overall, both centrals and satellites show similar gradients with stellar mass in both D$_n$4000 and H$\delta$ EW, but important differences in their environmental trends. Satellites show very significant gradients with changing over-density, and their stellar mass correlates the strongest with environmental over-density. At a fixed stellar mass, satellites in higher densities have larger D$_n$4000 and lower H$\delta$ EWs.

For centrals, there is little environmental dependency at the lowest masses (where the increase of D$_n$4000 and the decrease of H$\delta$ EW is purely stellar-mass driven), but above $10^{10.5}$\,M$_{\odot}$ even centrals show environmental dependencies. For a fixed stellar mass above $10^{10.5}$\,M$_{\odot}$, centrals residing in higher density regions are consistent with being older (higher D$_n$4000) and less star-forming (lower H$\delta$ EW) than their counter-parts residing in low-density regions by about 0.1 to 0.2\,dex (see also Tables~\ref{table_D4000_Mass_environment} and  \ref{table_Hdelta_Mass_environment}).

\subsection{D$_n$4000 and H$\delta$ for star-forming centrals \& satellites}

Figure \ref{D4000_Hd_vs_mass_Env_SFCentrals_SATS} shows how D$_n$4000 and H$\delta$ depend on environmental over-density and stellar mass for star-forming centrals \& satellites. Contrarily to the trends seen for the full population of galaxies, star-forming galaxies show D$_n$4000 and H$\delta$ EWs, which are almost insensitive to their environment when controlling for stellar mass (Tables~\ref{table_D4000_Mass_environment} and  \ref{table_Hdelta_Mass_environment}). This is even the case if they are satellites, where the increase of D$_n$4000 and the decrease of H$\delta$ EWs is purely driven by stellar mass. For a fixed stellar mass, star-forming galaxies show the same typical D$_n$4000 and H$\delta$ EW regardless of their environment: the gradients in D$_n$4000 and H$\delta$ are essentially all along the horizontal direction in Figure \ref{D4000_Hd_vs_mass_Env_SFCentrals_SATS}. Furthermore, while star-forming satellites reside in typically higher over-densities, the overall D$_n$4000 and H$\delta$ properties of satellites and central SFGs are, to first order, indistinguishable from those of star-forming centrals. The similarity in trends between star-forming satellites and centrals could point towards star-forming satellites having mostly been recently accreted and thus not having been satellites for long enough to lead to any noticeable difference.

\subsection{D$_n$4000 and H$\delta$ for quiescent centrals \& satellites}

Quiescent galaxies reveal statistically significant environmental dependencies of D$_n$4000 and H$\delta$ EW when controlling for stellar mass, regardless of being satellites or centrals (Figure \ref{D4000_Hd_vs_mass_Env_Qcentrals_vs_Sats} and Tables~\ref{table_D4000_Mass_environment} and \ref{table_Hdelta_Mass_environment}). There are important trends with stellar mass when controlling for environment for both quiescent centrals and quiescent satellites, but the mass dependence of H$\delta$ when controlling for environment is particularly weak (see Table \ref{table_Hdelta_Mass_environment}). Our results thus reveal that the environmental dependencies of both D$_n$4000 and H$\delta$ are fully driven by the quiescent population. In addition, as Figure \ref{D4000_Hd_vs_mass_Env_Qcentrals_vs_Sats} shows, quiescent centrals reveal the greatest correlation between their stellar mass and their environmental density. This suggests a stronger connection between their growth and their environment, likely enabling growth through various minor mergers and also an earlier formation time for higher density centrals.

\section{Discussion}\label{Discussion}

We have shown that there are statistically significant gradients (albeit small in absolute terms) in typical D$_n$4000 and H$\delta$ EW for the full massive galaxy population at $z\sim0.6-1.0$, in the mass-environment 2D space (Tables \ref{table_D4000_Mass_environment} and \ref{table_Hdelta_Mass_environment}). While these are statistically significant, only a survey like LEGA-C can fully reveal them in detail, going significantly beyond previous studies \citep[e.g.][]{PA2020}. This is due to the LEGA-C's high S/N, typical low individual uncertainties and the large samples and wide range of parameter space studies with a single data-set.

We find that while the primary predictor of D$_n$4000 and H$\delta$ EW at $z\sim1$ for massive galaxies is stellar mass \citep[in agreement with previous studies, e.g.][]{Muzzin12}, there are statistically significant environmental trends, even when controlling for stellar mass (Tables~\ref{table_D4000_Mass_environment} and \ref{table_Hdelta_Mass_environment}). By splitting the sample into centrals and satellites, and in terms of star-forming galaxies and quiescent galaxies, we reveal that the environmental trends of D$_n$4000 and H$\delta$ EW are exclusively seen in and driven by $UVJ$ quiescent galaxies. Star-forming galaxies reveal D$_n$4000 and H$\delta$ EWs, which depend solely on their stellar mass and are completely independent of the environment within the studied sample. The strong environmental trends seen for satellite galaxies are fully driven by the trends that hold only for quiescent galaxies, combined with the strong environmental dependency of the quiescent fraction \citep[e.g.][]{Peng10,Muzzin12,PA2019,Wang2021}.

For quiescent galaxies, our results can be interpreted as revealing an earlier assembly history in higher densities/higher mass dark matter haloes at a fixed stellar mass. The strong correlation between stellar mass and environmental density of quiescent centrals also suggests a relation between the environment and the star-formation history, including the potential of growth by multiple minor mergers. As quiescent satellites reveal the same patterns and become more dominant in higher density regions, growing by minor mergers in high density regions may allow to crystallise/strengthen the gradients with environment, while also tightening the relation between stellar mass and environmental over-density for quiescent centrals.

The results for the quiescent population imply an important assembly bias: galaxies in higher-density regions seem to be typically older and might have formed earlier than the low-density counterparts, in line with several results in the literature \citep[e.g.][]{Coop2010,Barbera2010,Barbera2014,McDermid2015,Gallazzi2021}. This is qualitatively consistent with various state-of-the-art cosmological simulations \citep[e.g.][]{Matthee2017EAGLE,Wechsler2018,Martizzi2020}.

However, as our results show, star-forming galaxies reveal no signature of the environment in their D$_n$4000 and H$\delta$ indices. This suggests that the light-weighted star-formation history revealed by these indicators is completely independent of the environment\footnote{Note that global smoothing width used by \cite{Darvish2017} is 0.5 Mpc \citep[see][for full details]{Darvish15b,Darvish2016,Darvish2017}, which corresponds to the typical virial radius for X-ray groups and clusters in the COSMOS field \citep[][]{Finoguenov2007,George2011}.}. This is the case regardless of these galaxies being satellites or centrals at $z\sim0.6-1.0$.

These results are also fully consistent with $z\sim0.8$ surveys that focused on star-forming galaxies selected using H$\alpha$ and using photometry to estimate H$\alpha$ emitters' D$_n$4000 \citep[][]{Sobral11}. In agreement with previous results, we find that changes in D$_n$4000 are fully driven by stellar mass and there is no dependence on the environment, for star-forming galaxies. Spectroscopic surveys at $z\sim0.8$ using VIMOS, but not going as deep as LEGA-C (and thus relying on stacking), also found similar results \citep[see][]{PA2020}.

The strong difference between how D$_n$4000 and H$\delta$ indices depend on stellar mass and environment for star-forming galaxies and quiescent galaxies can likely be explained by different physical and/or observational effects. Observationally, trends with environment (which are subtle) for star-forming galaxies may be hidden due to outshining from the most recent population of stars. By definition, star-forming galaxies will have high enough specific star formation rates (sSFR) to have a high SFR in respect to their stellar mass. This makes it easier to hide different past star-formation histories and to only see a population more directly linked with the current SFR. Several studies reveal the star-formation ``main sequence" is independent of environment, and thus SFR will depend only on mass \citep[e.g.][]{Koyama13}, making it easier to observationally hide different past star-formation histories. Recent results from the SDSS survey \citep[e.g.][]{Gallazzi2021} reveal that there are small differences on the stellar populations as a function of environment, particularly the age (including mass-weighted age) and metallicity of massive star-forming galaxies. Such results seem to point to an environmental effect on the old component of the galaxies, rather than on the recent SF \citep[][]{Gallazzi2021}.

A more physical interpretation is related to how star-forming galaxies become quenched and how the environment might be driving it. If the quenching mechanism were to be slow, one would expect to see trends in at least H$\delta$ with environment, at fixed mass. LEGA-C is certainly sensitive enough and it provides a large enough sample and a large enough range to see subtle effects. That is nonetheless not seen for star-forming galaxies, giving strength to relatively ``quick" environmental quenching mechanisms, in agreement with previous results, and the so-called ``light-switch" quenching \citep[e.g.][]{Muzzin12, Darvish2016}. Furthermore, \cite{MatSch2019} shows that the scatter in the ``main sequence" of star-forming galaxies is not really affected by star-forming satellites. Future studies can also try to investigate any evidence for strangulation and slightly elevated gas metallicities in higher density/satellite star-forming galaxies in respect to lower density and central counterparts.

Overall, our results reveal significant D$_n$4000 and H$\delta$ trends with both stellar mass and environment already in place at $z\sim0.8$ for both centrals and satellites. We reveal environmental effects beyond simply setting the quenched fraction. The environment clearly leaves a legacy in the stellar populations of quiescent galaxies, but a survey like LEGA-C is required to reveal it. The results therefore suggest an earlier assembly time and an older age at high densities for a fixed stellar mass.

\section{Conclusions}\label{conclusion}

We have explored the large LEGA-C sample (DR3; \citealt{vanderWel2021}) to investigate the role of the environment and stellar mass on galaxy evolution at $z\sim0.6-1$ in the COSMOS field. We show that LEGA-C samples the full range of environments without strong biases in COSMOS at $z\sim0.6-1$, from the low-density field to rich clusters. We explore the uniquely deep spectra to measure D$_n$4000 and H$\delta$ indices which provide crucial information on the recent and old stellar populations. We find that:
\begin{itemize}

\item The quiescent fraction in LEGA-C depends strongly on stellar mass and on environmental density, in agreement with previous studies, being close to $\approx2\%$ at $\approx10^{10.5}$\,M$_\odot$ in under-dense regions and $\approx100$\% at $\sim10^{12}$\,M$_\odot$ in high-density regions (Figure \ref{Q_fraction_full_sample}).

\item D$_n$4000 and H$\delta$ indices depend most strongly on stellar mass at $z\sim0.6-1$, but there are statistically significant environmental gradients (Tables \ref{table_D4000_Mass_environment} and \ref{table_Hdelta_Mass_environment}). At fixed stellar mass, massive galaxies at $z\sim0.6-1$ have higher D$_n$4000 and lower H$\delta$ in higher density regions when compared to their lower over-densities' counterparts (Figure \ref{D4000_Hd_vs_mass_Env}).

\item D$_n$4000 and H$\delta$ indices for star-forming galaxies depend strongly on stellar mass and we find no significant dependence on environment (Figure \ref{D4000_Hd_vs_mass_Env_SF_Q}). At fixed stellar mass, star-forming centrals and satellites have the same typical D$_n$4000 and H$\delta$ indices regardless of them residing in under- or over-densities (Figure \ref{D4000_Hd_vs_mass_Env_SFCentrals_SATS}).

\item D$_n$4000 and H$\delta$ indices for quiescent galaxies vary with both stellar mass and over-density (Figure \ref{D4000_Hd_vs_mass_Env_SF_Q} and Tables \ref{table_D4000_Mass_environment} and \ref{table_Hdelta_Mass_environment}). The environmental dependence at fixed stellar mass holds for both quiescent centrals and satellites, but quiescent centrals show the strongest relations with their environment. At a fixed stellar mass, quiescent galaxies, regardless of being satellites or centrals, have higher D$_n$4000 and lower H$\delta$ indices as a function of increasing over-density (Figure \ref{D4000_Hd_vs_mass_Env_Qcentrals_vs_Sats}). This suggests that quiescent galaxies residing in higher over-densities are older, formed earlier and/or quenched earlier.

\end{itemize}

These results show, for the first time, important environmental dependencies of D$_n$4000 and H$\delta$ at fixed stellar mass which are not just driven by the quiescent fraction rising with increasing environmental over-density. While we show that star-forming galaxies seem to have recent star-formation histories that are fully independent of their environment, quiescent galaxies reveal important environmental trends. These can be explored in the future with more detailed modelling of the full LEGA-C DR3 spectra, unveiling their star formation histories \citep[e.g.][]{Chauke2018}, metallicities, comparing the results with those in the local Universe \citep[e.g.][]{Gallazzi2021} and with predictions from simulations \citep[e.g.][]{Martizzi2020}.

\

\acknowledgments

We thank the reviewer for several valuable comments which improved the clarity of the manuscript. P.F.W. acknowledges the support of the fellowship by the East Asian Core Observatories Association. This work is based on observations made with ESO VLT Telescopes at the La Silla Paranal Observatory under programmes ID 194-A.2005 and 1100.A-0949 (The LEGA-C Public Spectroscopy Survey). This project has received funding from the European Research Council (ERC) under the European Union -- Horizon 2020 research and innovation program (grant agreement No. 683184).

\bibliography{bib}

\end{document}